\documentclass[aps,prd,superscriptaddress,nofootinbib,tighten,preprint]{revtex4}
\pdfoutput=1
\usepackage[utf8]{inputenc}
\usepackage[english]{babel}
\usepackage{amsmath}
\usepackage{subfigure}
\usepackage{amsfonts}
\usepackage{amssymb}
\usepackage{color}
\usepackage{cancel}
\usepackage{graphicx}
\newcommand{\be}{\begin{equation}}
\newcommand{\ee}{\end{equation}}
\newcommand{\bea}{\begin{eqnarray}}
\newcommand{\eea}{\end{eqnarray}}

\newcommand{\nn}{\nonumber}

\catcode`@=12

\preprint{IP/BBSR/2017-15}
\preprint{HIP-2017-34/TH} 

\begin{document}
\title{Singlet-Triplet Fermionic Dark Matter and LHC Phenomenology}

\author{Sandhya Choubey}
\email{sandhya@hri.res.in}
\affiliation{Harish-Chandra Research Institute, Chhatnag Road,
Jhunsi, Allahabad 211 019, India}
\affiliation{Homi Bhabha National Institute,
Training School Complex, Anushaktinagar, Mumbai - 400094, India}
\affiliation{Department of Physics, School of
Engineering Sciences, KTH Royal Institute of Technology, AlbaNova
University Center, 106 91 Stockholm, Sweden}
\author{Sarif Khan}
\email{sarifkhan@hri.res.in}
\affiliation{Harish-Chandra Research Institute, Chhatnag Road,
Jhunsi, Allahabad 211 019, India}
\affiliation{Homi Bhabha National Institute,
Training School Complex, Anushaktinagar, Mumbai - 400094, India}
\author{Manimala Mitra}
\email{manimala@iopb.res.in}
\affiliation{Institue of Physics, Sachivalaya Marg, Bhubaneswar,
Odisha 751005, India}
\affiliation{Homi Bhabha National Institute,
Training School Complex, Anushaktinagar, Mumbai - 400094, India}
\author{Subhadeep Mondal}
\email{subhadeep.mondal@helsinki.fi}
\affiliation{Department of Physics, Helsinki Institute of Physics,
P. O. Box 64, FI-00014, University of Helsinki, Finland}

\begin{abstract} 
It is well known that for the pure standard model triplet fermionic WIMP-type dark matter (DM),
the relic density is satisfied around 2 TeV. For such a heavy mass particle, the production cross-section at 13 TeV run of LHC will be  very small.
Extending the model further with a singlet fermion and a triplet scalar, DM relic density can be satisfied for even much lower masses. The lower mass DM can be copiously produced at LHC and hence the model can be tested at collider. For the present model we
have studied the multi jet ($\geq 2\,j$) + missing energy ($\cancel{E}_{T}$) signal and show that this can be 
detected in the near future of the LHC 13 TeV run. 
We also predict that the present model is testable
by the earth based DM direct detection experiments like Xenon-1T and in future by Darwin. 
\end{abstract}
\maketitle
\newpage 
\section{Introduction}
The Standard Model (SM) of elementary particles is a very well established and successful theory. 
With the discovery of the Higgs boson at the LHC, the last missing piece of the SM has been found. 
So far, all observations at the collider experiments are reasonably consistent with the SM 
cementing its position  even further. However, despite this success story  it is well accepted that the 
SM is not the full theory of nature. Rather, SM is widely looked as a low energy effective limit of a more complete
underlying theory. The reasons to believe that SM needs to be extended are both theoretical as well as 
observational. Amongst the most compelling observational evidences of physics Beyond the SM (BSM) is the issue 
of the Dark Matter (DM). The DM, if it is a particle, should be massive, chargeless and weakly interacting such that 
its relic abundance should be consistent with the observational data on DM. The 
SM fails to provide any such candidate. The only weakly interacting chargeless particle in the SM is the neutrino and 
it is postulated to be massless. Of course experimental data have now given conclusive evidence that neutrinos 
are massive - which is another compelling reason to extend the SM to accommodate neutrino mass. However 
neutrinos, even if massive in a BSM theory, can only be hot dark matter candidate and it is well known that hot dark matter is inconsistent with structure formation of the universe. 
Therefore, we need a 
BSM theory that can provide either cold or warm dark matter candidate. 

Presence of DM is a very well established fact which is supported by many
evidences. The first evidence of DM came from the observation of 
the flatness of the rotation curve of the Coma cluster 
by a Swedish scientist, Zwicky, in 1930 \cite{Sofue:2000jx}. Other strong evidences that 
support the DM theory include gravitational
lensing \cite{Bartelmann:1999yn}, recent observation of bullet
cluster by NASA's Chandra satellite \cite{Clowe:2003tk, Harvey:2015hha},
and from the measurement of Cosmic Microwave Background (CMB) radiation
\cite{Hinshaw:2012aka, Ade:2015xua}. Planck \cite{Hinshaw:2012aka} and
WMAP \cite{Ade:2015xua} have precisely measured
the amount of DM present in the universe and have given a $2\sigma$ bound on the DM
relic density as, 
\begin{eqnarray}
\Omega h^{2} = 0.1199 \pm 0.0027
\label{dm-bound}
\end{eqnarray}     
Many different classes of DM candidates like the Weakly Interacting Massive Particle (WIMP),
Strongly Interacting Massive Particle (SIMP) and Feebly Interacting Massive Particle (FIMP), 
have been proposed in the literature. 
Each type can solve the DM puzzle in its unique way.  In this work we will consider a model with 
a WIMP-type fermionic DM candidate by extending the SM particle spectrum and study in detail the DM phenomenology. 
We will also study the collider signal of the DM at the 13 TeV run of the 
LHC and its effect will basically manifest as the missing energy associated with hard jets.
In \cite{Hisano:2002fk,Hisano:2003ec,Hisano:2004ds,Cirelli:2005uq,Hisano:2006nn, Ma:2008cu,
Cohen:2013ama}, a model has been proposed where 
the fermionic DM belongs to the triplet representation of the SM. 
An extra discrete $\mathbb{Z}_{2}$
symmetry stabilizes the DM making the neutral part of the triplet fermion
as the viable candidate for the DM.
If the SM is extended by only the triplet fermion \cite{Cirelli:2005uq,Ma:2008cu}, then the 
main co-annihilation processes take part in the DM relic density calculation are 
mediated by the charged gauge boson $W^{\pm}$, and for this case 
the correct value of DM relic density is obtained for DM mass around 2.3 TeV
\cite{Cirelli:2005uq, Ma:2008cu}. 
This high mass makes it difficult to produce the purely triplet fermionic DM at the 13 TeV 
collider, and hence to test this model at the LHC. Of course with a higher energy 
collider one might be able to to produce these heavy fermions. 
Another major drawback of such high mass DM is that when the DM annihilate to
gamma rays via $W^{\pm}$-mediated one loop diagrams, then
for such high DM mass, the annihilation
is increased by the Sommerfeld enhancement (SE) factor
\cite{Hisano:2003ec,Hisano:2004ds,Cirelli:2005uq}. This is ruled out from
the indirect search of the HESS data \cite{Abramowski:2013ax}.
In this paper, we propose an extension of the SM that accommodates both high as well as low mass fermionic DM such that it can be produced and tested 
at the 13 TeV run of the LHC. The low mass DM regime do not have any significant SE 
enhancement (because the DM mass becomes comparable to the mediator
mass inside the loops) and hence are safe from the
gamma ray indirect detection bounds put by the Fermi-LAT collaboration \cite{Ackermann:2015lka}. 
Our proposed extension of the particle content includes 
one SM singlet fermion and  SM triplet fermion
\cite{Chardonnet:1993wd, Biggio:2011ja, Hirsch:2013ola,
Chaudhuri:2015pna, Bhattacharya:2017sml}.
The scalar sector is also 
extended to include a SM triplet scalar. The $\mathbb{Z}_{2}$ charge of these BSM particles 
is arranged in such a way that there is a mixing between the neutral component of the 
triplet fermion and the singlet fermion, that generates two mass eigenstates
for the neutral fermions. The lower mass eigenstate becomes the viable DM candidate.
The neutral and charged components of the SM doublet and triplet scalars also mix, that gives rise to two physical neutral Higgs scalars and one charged Higgs scalar. The presence of these extra scalars opens up additional annihilation and co-annihilation processes between the two DM candidates 
which effectively reduces the mass of the DM for which the current DM relic density bound can be easily satisfied.
For low mass DM we give the prediction for the annihilation
of the DM to two gamma rays by one loop process.
In addition, these lower mass DM fermions ($\sim$ 100 GeV) can be observed with large 
production cross-section at the 13 TeV LHC. We perform a detailed collider phenomenology 
of the DM model. In this work we will 
consider multi jets + missing energy signal in the final state for searching the DM.
We study in detail the dominant backgrounds for such type of 
signal. The SM backgrounds are reduced by applying suitable cuts that  
increases the statistical significance of detection for  the fermionic 
DM with the low luminosity run of the LHC.
A final comment is in order. 
It is possible to embed our model in a SO(10) GUT where the SU(2) triplet 
would belong to the 45 representation of SO(10) and would help 
in the gauge coupling unification, as was 
shown in \cite{Frigerio:2009wf, Mambrini:2016dca}. 

Rest of the manuscript is organised as follows. In section \ref{onlytriplet}
we briefly discuss the triplet fermionic 
DM model proposed in \cite{Cirelli:2005uq,Ma:2008cu} and its corresponding DM constraints. 
In section \ref{sing-trip} we details of our proposed model. In section \ref{constraints}
we list down all the constraints imposed on the DM model parameter space 
from the existing data. In section \ref{dmresults} we present our main results on 
the DM phenomenology of our model. 
In section \ref{indirect-detection}, we show the predicted gamma ray flux from our 
model and compare it against the bounds from the Fermi-LAT data.
In section \ref{lhc} we study in detail the collider 
phenomenology of our model for the 13 TeV LHC. Finally, we summarise our results 
in section \ref{conclusion}.

\section{Triplet Fermionic Dark Matter \label{onlytriplet}}

\def\I{i}
\begin{center}
\begin{table}[h!]
\begin{tabular}{||c|c|c|c||}
\hline
\hline
\begin{tabular}{c}
    Gauge\\
    Group\\ 
    \hline
    $SU(3)_{c}$\\ 
    \hline
    $SU(2)_{L}$\\ 
    \hline
    $U(1)_{Y}$\\ 
    \hline
    $\mathbb{Z}_{2}$\\
    
\end{tabular}
&

\begin{tabular}{c|c|c}
    \multicolumn{3}{c}{Baryon Fields}\\ 
    \hline
    $Q_{L}^{i}=(u_{L}^{i},d_{L}^{i})^{T}$&$u_{R}^{i}$&$d_{R}^{i}$\\ 
    \hline
    $3$&$3$&$3$\\
    \hline
    $2$&$1$&$1$\\ 
    \hline
    $1/6$&$2/3$&$-1/3$\\
    \hline
    $+$&$+$&$+$\\ 
     
\end{tabular}
&
\begin{tabular}{c|c|c}
    \multicolumn{3}{c}{Lepton Fields}\\
    \hline
    $L_{L}^{i}=(\nu_{L}^{i},e_{L}^{i})^{T}$ & $e_{R}^{i}$ &~ $\rho$~ \\
    \hline
    $1$&$1$&$1$\\
    \hline
    $2$&$1$&$3$\\
    \hline
    $-1/2$&$-1$&$0$\\
    \hline
    $+$&$+$&$-$\\
    
\end{tabular}
&
\begin{tabular}{c}
    \multicolumn{1}{c}{Scalar Fields}\\
    \hline
    $\phi_{h}$\\
     \hline
    $1$\\
    \hline
    $2$\\
    \hline
    $1/2$\\
    \hline
    $+$\\
     
\end{tabular}\\
\hline
\hline
\end{tabular}
\caption{\label{tab:modelMa} Particle content and their corresponding
charges under various symmetry groups.}
\label{tab1}
\end{table}
\end{center}

\begin{figure}[h!]
	\centering
	\includegraphics[width=0.45\textwidth]{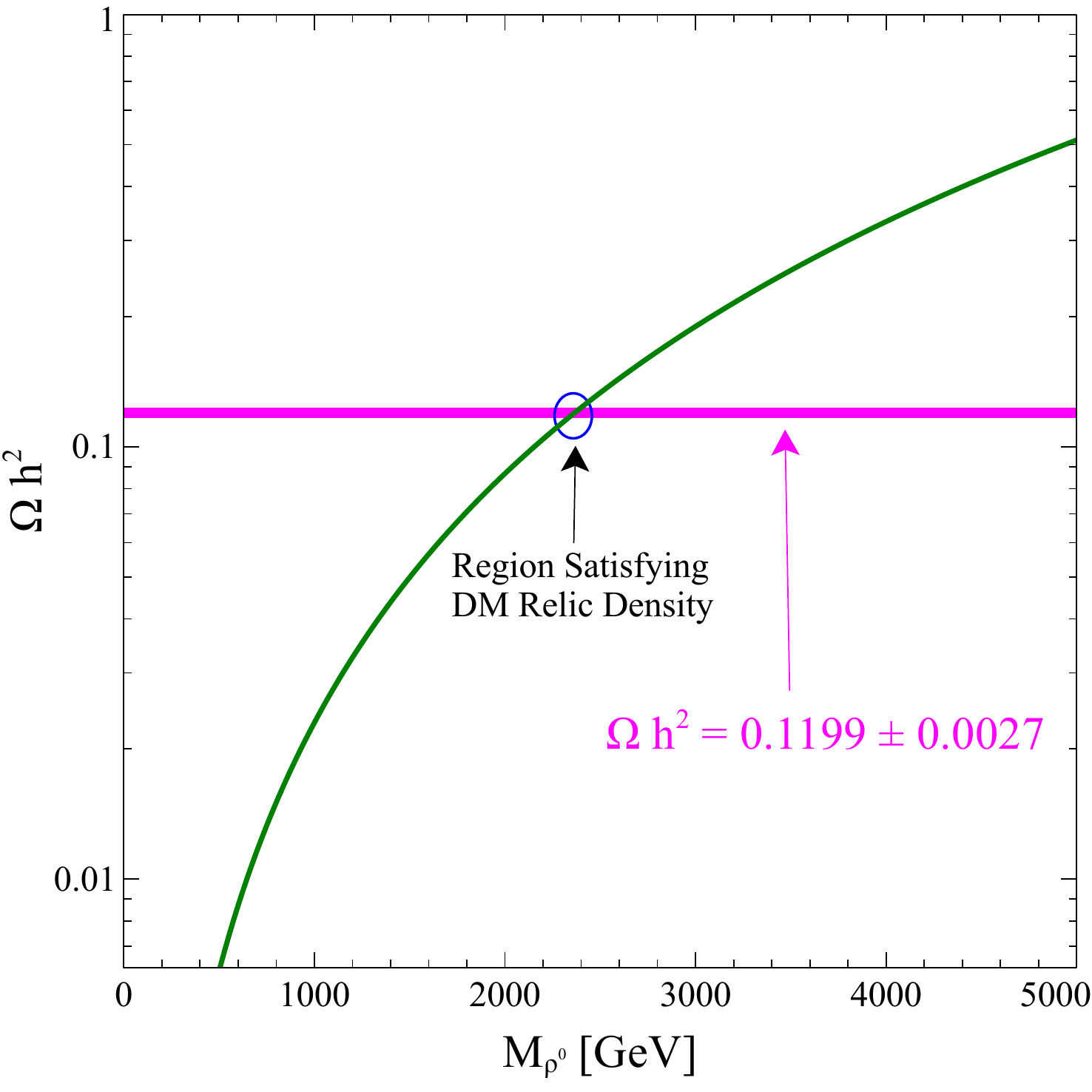}
	\caption{Variation of relic density $\Omega h^{2}$ with the mass of the triplet
	DM $M_{\rho^{0}_{1}}$.}
	\label{1a}
\end{figure}

In this case the SM particle content is extended with just a left handed fermionic triplet field $\rho$ \cite{Cirelli:2005uq, Ma:2008cu}. There is an additional $\mathbb{Z}_{2}$ symmetry imposed on the model such that the  triplet is odd under it, while all SM particles are even under this symmetry. The particle content of the model and their charges under the symmetries of the model is given in Table \ref{tab:modelMa}. The $\mathbb{Z}_{2}$ symmetry forbids all the Yukawa couplings of $\rho$ with the SM fermions and the complete Lagrangian includes just the additional kinetic energy term for the triplet ($\mathcal{L}_{\rho}$) along with the SM 
Lagrangian ($\mathcal{L_{SM}}$),
\begin{eqnarray}
\mathcal{L} = \mathcal{L_{SM}} + \mathcal{L}_{\rho}\,.
\end{eqnarray}
The Lagrangian for triplet field $\rho$ takes the following form,
\begin{eqnarray}
\mathcal{L}_{\rho} = Tr[\bar{\rho}\, i \gamma^{\mu} D_{\mu}\, \rho]
\,,
\end{eqnarray} 
where the covariant derivative $D_{\mu}$ takes the following form,
\begin{eqnarray}
D_{\mu} = \partial_{\mu} - i g\, T^{adj}_{i}\, W_{i}\,,
\end{eqnarray}
\begin{figure}[h!]
	\centering
	\includegraphics[width=0.75\textwidth]{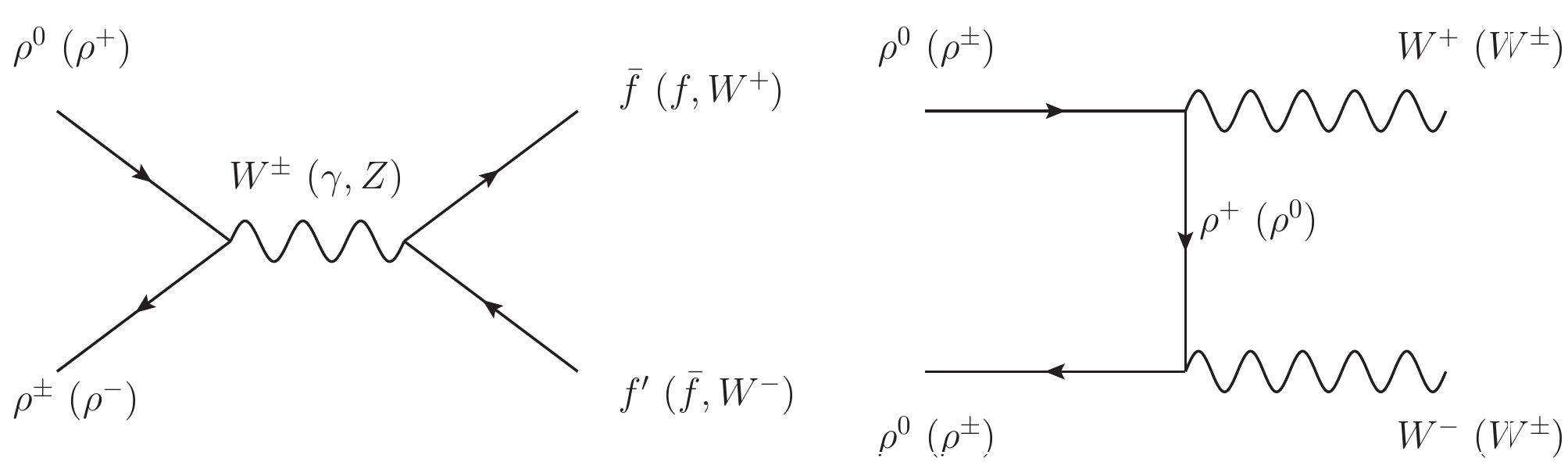}
	\caption{Pure triplet fermions DM annihilation and co-annihilation diagrams.}
	\label{pure-trip}
\end{figure}
where $g$ and $W_{i}$ are the $SU(2)_{L}$ gauge coupling and gauge field, respectively, and $T^{adj}_{i}$'s are the
$SU(2)_{L}$ generators in the adjoint representation. The $\mathbb{Z}_{2}$ symmetry makes $\rho^0$, the chargeless component of $\rho$ stable and it becomes the DM. The annihilation and co-annihilation of the DM $\rho_0$, and $\rho^{\pm}$ proceed through SM gauge bosons, as shown in Fig.~\ref{pure-trip}. In Fig.\,\ref{1a} we show the $\rho^0$ relic abundance as a function of its mass $M_{\rho^0}$. 
From the figure one can notice that with the increase of the 
DM mass, its relic density also increases. This is because in the present case the velocity
times the DM annihilation
and co-annihilation cross sections vary inversely with the square of the DM mass. 
Hence, as the DM mass is increased, the DM annihilation cross-section decreases and as a result the DM relic density increases as it is inversely proportional to the velocity times cross section. From the figure we note 
 that the present day observed value of the DM relic density is satisfied around $M_{\rho^{0}} \sim 2370$ GeV. This has been also pointed out before in \cite{Ma:2008cu}. 
 
 Note that, while the model can be tested in direct and indirect detection experiments, due  to its heavy mass it is difficult to produce this DM candidate
at the 13 TeV or 14 TeV LHC search. One will need a very high energy collider to test this DM model. Minimal extension of the model by adding a gauge singlet fermion and a triplet scalar opens up the possibility to test the model at collider. Below, we discuss in detail the required extensions and the model predictions.

\section{Singlet Triplet Mixing \label{sing-trip}}

\def\I{i}
\begin{center}
\begin{table}[h!]
\begin{tabular}{||c|c|c|c||}
\hline
\hline
\begin{tabular}{c}
    Gauge\\
    Group\\ 
    \hline
    $SU(3)_{c}$\\ 
    \hline
    $SU(2)_{L}$\\ 
    \hline
    $U(1)_{Y}$\\ 
    \hline
    $\mathbb{Z}_{2}$\\     
\end{tabular}
&

\begin{tabular}{c|c|c}
    \multicolumn{3}{c}{Baryon Fields}\\ 
    \hline
    $Q_{L}^{i}=(u_{L}^{i},d_{L}^{i})^{T}$&$u_{R}^{i}$&$d_{R}^{i}$\\
    \hline
    $3$&$3$&$3$\\  
    \hline
    $2$&$1$&$1$\\ 
    \hline
    $1/6$&$2/3$&$-1/3$\\
    \hline
    $+$&$+$&$+$\\ 
     
\end{tabular}
&
\begin{tabular}{c|c|c|c}
    \multicolumn{4}{c}{Lepton Fields}\\
    \hline
    $L_{L}^{i}=(\nu_{L}^{i},e_{L}^{i})^{T}$ & ~$e_{R}^{i}$~ & ~$N^{\prime}$~ & ~$\rho$~ \\
    \hline
    $1$&$1$&$1$&$1$\\
    \hline
    $2$&$1$&$1$&$3$\\
    \hline
    $-1/2$&$-1$&$0$&$0$\\
    \hline
    $+$&$+$&$-$&$-$\\
    
\end{tabular}
&
\begin{tabular}{c|c}
    \multicolumn{2}{c}{Scalar Fields}\\
    \hline
    $\phi_{h}$&$\Delta$\\
    \hline
    ~~$1$~~&$1$\\
    \hline
    ~~$2$~~&$3$\\
    \hline
    ~~$1/2$~~&$0$\\
    \hline
    ~~$+$~~&$+$\\
     
\end{tabular}\\
\hline
\hline
\end{tabular}
\caption{Particle content and their corresponding
charges under various symmetry groups.} 
\label{tab:tab2}
\end{table}
\end{center}  


In this section, we present a minimal extension of the model, such that the mass of the DM can be suitably reduced and it can be produced at the LHC. To that end, we add an extra gauge singlet fermion which is also odd under the $\mathbb{Z}_{2}$ and an additional real triplet Higgs ($Y=0$). The particle content of our model and their respective charges are displayed in the Table \ref{tab:tab2}.
The corresponding Lagrangian is given by,
\begin{eqnarray}
\mathcal{L} &=& \mathcal{L}_{SM} + Tr \left[ \bar{\rho}\,i\, \gamma^{\mu} D_{\mu} \rho \right]
+ \bar{N^{\prime}}\,i\, \gamma^{\mu} D_{\mu} N^{\prime}
+Tr[(D_{\mu}\Delta)^{\dagger} (D^{\mu} \Delta)] -V(\phi_{h}, \Delta) \nn \\
&& - Y_{\rho \Delta}\, (Tr[\bar{\rho}\, \Delta]\, N^{\prime} + h.c.) 
- M_{\rho}\, Tr[\bar{\rho^{c}} \rho]
 - M_{N^{\prime}}\, \bar{N^{\prime c}} N^{\prime} 
\label{mix-lag}
\end{eqnarray}
where the triplet fermion takes the following form,
\begin{eqnarray}
\rho=
\begin{pmatrix}
\frac{\rho_{0}}{2} & \frac{\rho^{+}}{\sqrt{2}} \\
\frac{\rho^{-}}{\sqrt{2}} & -\frac{\rho_{0}}{2}
\end{pmatrix}\,\,.
\label{phih}
\end{eqnarray}
The complete form of the potential $V(\phi_{h}, \Omega)$
takes the following form,
\begin{eqnarray}
V(\phi_{h}, \Delta) &=& -\mu_{h}^{2} \phi_{h}^{\dagger} \phi_{h}
+ \frac{\lambda_{h}}{4} (\phi_{h}^{\dagger} \phi_{h})^{2}
+\mu_{\Delta}^{2} Tr[\Delta^{\dagger} \Delta] + \lambda_{\Delta} (\Delta^{\dagger} \Delta)^{2}
+ \lambda_{1}\, (\phi_{h}^{\dagger} \phi_{h})\, {\rm Tr}\,[\Delta^{\dagger} \Delta] \nonumber \\
&&+ \lambda_{2}\,\left(Tr[\Delta^{\dagger} \Delta]\right)^{2} 
+ \lambda_{3}\,Tr[(\Delta^{\dagger} \Delta)^{2}]
+ \lambda_{4}\, \phi_{h}^{\dagger} \Delta \Delta^{\dagger} \phi_{h}
+ ( \mu \phi_{h}^{\dagger} \Delta \phi_{h} + h.c.)\,.
\end{eqnarray}
In general, one can also insert a term like $\phi_{h}^{\dagger} \Delta^{\dagger} \Delta \phi_{h}$,
but this term can be easily decomposed to two components
that give contribution to the terms with 
$\lambda_{1}$ and $\lambda_{4}$ couplings. Hence, we do not write this term separately in the potential.
We assume here that $\mu_{\Delta}^{2}$ is positive hence the
neutral component of the Higgs triplet will get small induced vev, because it
has coupling with the SM like Higgs, after electro
weak symmetry breaking (EWSB) which takes the following form,
\begin{eqnarray}
\langle \Delta_0 \rangle = v_{\Delta} = 
\frac{\mu v^{2}}{2 \left(\mu_{\Delta}^{2} + (\lambda_{4} + 2\lambda_{1})
\frac{v^{2}}{4} + (\lambda_3 + 2 \lambda_2) \frac{v_{\Delta}^2}{2}\right)}
\end{eqnarray} 
The Higgs doublet and real triplet take
the following form after taking the small fluctuation around the
vevs $v$ and $v_{\Delta}$, respectively,
\begin{eqnarray}
\phi_{h}=
\begin{pmatrix}
\phi^{+} \\
\dfrac{v+H +  i\, \xi}{\sqrt{2}}
\end{pmatrix}
\,\,\,\,\,\,\,\,\,\,\,\,
\Delta=
\begin{pmatrix}
\frac{\Delta_{0} + v_{\Delta}}{2}  & \frac{\Delta^{+}}{\sqrt{2}} \\
\frac{\Delta^{-}}{\sqrt{2}} & -\frac{\Delta_{0} + v_{\Delta}}{2} 
\end{pmatrix}\,\,.
\label{phih}
\end{eqnarray}
Since $\phi_{h}$ takes vev spontaneously which breaks the EWSB and $\Delta$ gets induced vev, we need to satisfy
the following criterion for the quadratic and quartic couplings,
\begin{eqnarray}
\mu_{h}^{2} > 0, \,\,\,\, \mu_{\Delta}^{2} > 0, \,\,\,\, \lambda_{h} > 0\,\,\,\, {\rm and}\,\,\,\, \lambda_{\Delta} > 0\,.
\end{eqnarray}
After symmetry breaking the $2 \times 2$ mass matrix for the CP even Higgs
scalars $H$ and $\Delta_{0}$ take the following form,
\begin{eqnarray}
M_{s}
= \frac{1}{2}
\begin{pmatrix}
\lambda\, v^{2} & v\, v_{\Delta} (2 \lambda_{1}  + \lambda_{4}) - 2\, \mu\, v \\
v\, v_{\Delta} (2 \lambda_{1}  + \lambda_{4}) - 2\,\mu\, v &
2\, v_{\Delta}^{2} (\lambda_{3} + 2 \lambda_{2}) + \frac{\mu\, v^{2}}
{v_{\Delta}}
\end{pmatrix}
\end{eqnarray}
After diagonalisation of the above matrix we will get the physical Higgses $h_1$ and $h_2$
with masses $M_{h_1}$ and $M_{h_2}$, respectively. If the mixing angle between $h_1$
and $h_2$ is $\alpha$, then the mass and flavor eigenstates can be written in
the following way,
\begin{eqnarray}
h_{1} &=& \cos \alpha\, H + \sin \alpha\, \Delta_{0} \nn \\ 
h_{2} &=& -\sin \alpha\, H + \cos \alpha\, \Delta_{0}
\end{eqnarray}
The CP odd field $\xi$ becomes
Goldstone boson which is ``eaten'' by the SM gauge boson $Z$. In addition to the
mixing between $H$ and $\Delta_{0}$, the charged scalars will also be mixed 
and one of them
will be the Goldstone boson ``eaten'' by 
$W^{\pm}$. We can write them in the
physical basis 
in the following way,
\begin{eqnarray}
G^{\pm} &=& \cos \delta \, \phi^{\pm} + \sin \delta \, \Delta^{\pm} \nn \\ 
H^{\pm} &=& -\sin \delta \, \phi^{\pm} + \cos \delta\, \Delta^{\pm}
\end{eqnarray}
where the mixing angle depends on the
strength of the vevs of doublet and
triplet, {\it  i.e.}, 
\begin{eqnarray}
\tan \delta = \frac{2\,v_{\Delta}}{v}
\,.
\label{eq:vdelta}
\end{eqnarray}
The quadratic and quartic couplings have the following form in terms
of the CP even Higgs masses $M_{h_1}$ and $M_{h_2}$, the mixing angle between them $\alpha$, the charge scalar mass and the mixing angle
between the charged scalars $\delta$: 
\begin{eqnarray}
\mu &=& \frac{M_{H^{\pm}}^{2}
	\sin \delta \cos \delta}{v}\,,\nn \\
\lambda_{3} + 2\,\lambda_{2} &=& \dfrac{ M_{h_{1}}^{2} + M_{h_{2}}^{2} +
	(M_{h_{2}}^{2} - M_{h_{1}}^{2})\cos 2 \alpha - 2\,M_{H^{\pm}}^{2} \cos^{2}\delta}{2\,v_{\Delta}^{2}} \,,\nn \\
\lambda_{h} &=& \dfrac{M_{h_{1}}^{2} + M_{h_{2}}^{2} +
	(M_{h_{1}}^{2} - M_{h_{2}}^{2})\cos 2 \alpha}{v^{2}},\nn\\
\lambda_{4} + 2\,\lambda_{1} &=& \dfrac{(M_{h_{1}}^{2}-M_{h_{2}}^{2})
	\sin 2 \alpha + M^{2}_{H^{\pm}} \sin 2\, \delta
}{v\, v_{\Delta}}\,,\nn\\
\mu_{h}^{2} &=& \lambda_{h}\, \frac{v^{2}}{4} +(\lambda_{4} + 2 \lambda_{1})\,\frac{v_{\Delta}^{2}}{4}
- \mu\, v_{\Delta}\,.
\label{constraints-eq}
\end{eqnarray}
The vev of the Higgs triplet is constrained by the data on 
the ratio $\frac{M_{W}^{2}}{\cos^{2} \theta_{w} M_{Z}^{2}}$, which limits 
$v_{\Delta} < 12$ GeV \cite{Erler:2004nh, Erler:2008ek}.
The value of $M_{h_2}$ needs to satisfy the perturbativity limit on the
quartic couplings which is $\lambda < 4 \pi$. The quartic couplings are also
bounded from the below \cite{Chen:2008jg} and as long as all the quartic
couplings are positive, we do not need to worry about the lower bounds.
From Eq.\,(\ref{constraints-eq}) we see that 
by choosing a suitable value for the free parameter $\mu$ which has mass dimension,
we can keep all the quartic couplings in the perturbative regime. 

In Eq.\,(\ref{mix-lag}), $Y_{\rho \Delta}$ is the Yukawa term relating the fermionic triplet
with the fermionic singlet. When the neutral component of $\Delta$ takes vev, the mass matrix for the
fermions takes the following form,
\begin{eqnarray}
M_{F}
= 
\begin{pmatrix}
M_{\rho} & \frac{Y_{\rho\Delta} v_{\Delta}}{2} \\
\frac{Y_{\rho\Delta} v_{\Delta}}{2} & M_{N^{\prime}} 
\end{pmatrix}
\,.
\end{eqnarray} 
The lightest component of the eigenvalues of this matrix will be the stable DM.
Relation between the the mass eigenstates and weak eigenstates are as follows:
\begin{eqnarray}
\rho_{2}^{0} &=& \cos\beta\, \rho_{0} + \sin\beta\, N^{\prime c} \nn \\ 
\rho_{1}^{0} &=& -\sin\beta\, \rho_{0} + \cos\beta\, N^{\prime c}
\end{eqnarray}
Therefore, the tree level mass eigenstates are,
\begin{eqnarray}
M_{\rho_{1}^{0}} &=& \frac{1}{2} \left(M_{\rho} + M_{N^{\prime}} - \sqrt{(M_{\rho} - M_{N^{\prime}})^{2} +
4\left(\frac{Y_{\rho\Delta} v_{\Delta}}{2}\right)^{2}} \right)\,, \nn \\
M_{\rho_{2}^{0}} &=& \frac{1}{2} \left(M_{\rho} + M_{N^{\prime}} + \sqrt{(M_{\rho} - M_{N^{\prime}})^{2} +
4\left(\frac{Y_{\rho\Delta} v_{\Delta}}{2}\right)^{2}} \right)\,, \nn \\
\tan 2 \beta &=& \frac{Y_{\rho \Delta} v_{\Delta}}{M_{\rho} - M_{N^{\prime}}}
\,.
\end{eqnarray}
In terms of $M_{\rho_{1}^{0}}$ and $M_{\rho_{2}^{0}}$ we can express the Yukawa coupling
$Y_{\rho \Delta}$ in the following way:
\begin{eqnarray}
Y_{\rho \Delta} &=& \frac{(M_{\rho_{2}^{0}} - M_{\rho_{1}^{0}}) \sin2\beta}{2\, v_{\Delta}}\,, \nn \\
&=& \frac{\Delta M_{21} \sin2\beta}{2 v_{\Delta}}
\label{mix-coup}
\end{eqnarray}
where $\Delta M_{21} = (M_{\rho_{2}^{0}} - M_{\rho_{1}^{0}})$ represents the mass difference between $M_{\rho_{2}^{0}}$ and $ M_{\rho_{1}^{0}} $. Therefore,
one can increase the Yukawa coupling $Y_{\rho\Delta}$  by
increasing the mass difference $\Delta M_{21}$ or the singlet
triplet fermionic mixing angle, or decreasing the triplet vev $v_{\Delta}$. We have kept the mass of
charged component ($\rho^{\pm}$) of triplet fermion equal
to the mass of $\rho_{2}^{0}$
with the mass gap of pion {\it i.e.} $M_{\rho^{\pm}} = M_{\rho_{2}^{0}} + 0.16$ GeV.

A further discussion is in order. 
For the present model we can generate the neutrino mass by Type I seesaw
mechanism just by introducing SM singlet right handed neutrinos.
In other 
variants of the triplet fermionic DM model, neutrino masses were generated
by using the Type III seesaw mechanism and radiatively
by the authors of \cite{Abada:2008ea} and \cite{Ma:2008cu,Chardonnet:1993wd,Biggio:2011ja,Hirsch:2013ola}, respectively.

\section{Constraints used in Dark Matter Study \label{constraints}}

Below, we discuss different constraints that we take into account. This includes the constraints from relic density, the direct detection constraints, as well as the invisible Higgs decay. 

\subsection{SI direct detection cross section} 

\begin{figure}[]
	\centering
	\includegraphics[angle=0,height=5.5cm,width=16.5cm]{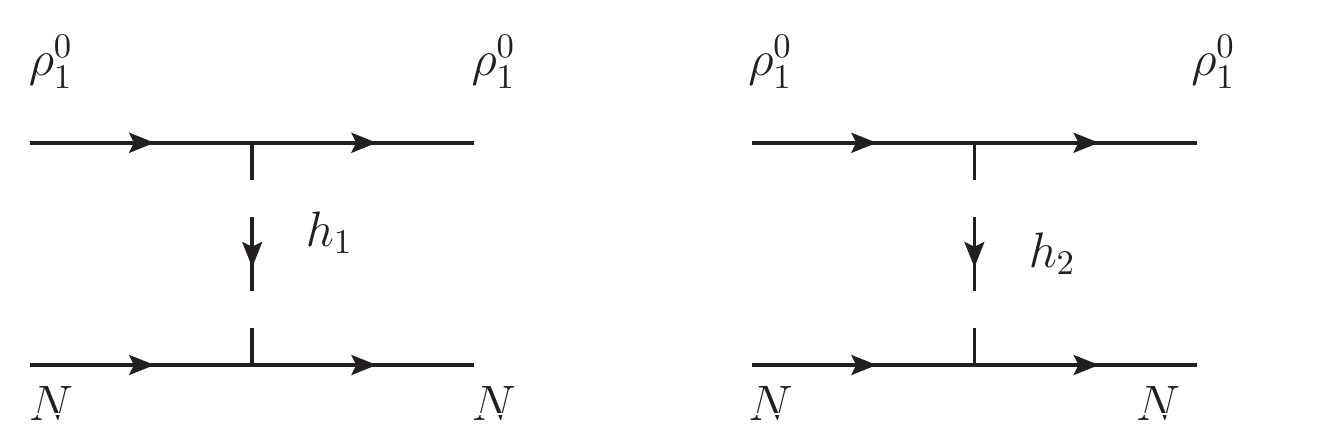}
    \caption{SI direct detection scattering processes between DM and
    nucleaon of the nucleus.}
	\label{4a}
\end{figure}

The Feynman diagrams in Fig.\,\ref{4a} show the spin independent (SI) direct detection
(DD) scattering processes between the DM and the nucleon ($\rho^{0}_1\, N \rightarrow \rho^{0}_1\, N$), which are mediated by the
two Higgses $h_1$ and $h_2$, respectively, through the t-channel process. Since DM interacts 
very weakly with the nucleon, one can safely calculate the
cross-section for this process in the $t \rightarrow 0$ limit, where $t$ is the Mandelstam variable corresponding to the square of the 
four-momentum transfer. The expression for
the above process takes the following form,
\begin{eqnarray}
\sigma_{SI} &=& \frac{\mu_{red}^{2}}{\pi}\,\left[ \frac{M_{N} f_{N}}{v}
\left(\frac{g_{\rho^{0}_{1} \rho^{0}_{1} h_{2}} \sin \alpha}{M_{h_2}^{2}} -
\frac{g_{\rho^{0}_{1} \rho^{0}_{1} h_{1}} \cos \alpha}{M_{h_1}^{2}} \right) \right]^{2}
\label{eq:si}
\end{eqnarray}
where the quantity  $f_N$  is the nucleon form factor and it is equal to
$0.3$ \cite{Belanger:2008sj} while 
$\mu_{red}$ is the reduced mass between the 
DM mass ($M_{\rho^{0}_{1}}$) and the nucleon mass ($M_{N}$) and is given by
\begin{eqnarray}
\mu_{red} &=& \frac{M_{N} M_{\rho^{0}_{1}}}{M_{N} + M_{\rho^{0}_{1}}}\,, 
\label{red}
\end{eqnarray}
The  couplings in Eq.~(\ref{eq:si}) $g_{\rho^{0}_{1} \rho^{0}_{1} h_{1}}$  and 
$g_{\rho^{0}_{1} \rho^{0}_{1} h_{2}}$  are given by,
\begin{eqnarray}
g_{\rho^{0}_{1} \rho^{0}_{1} h_{1}} &=& \frac{Y_{\rho\Delta}}{2} \sin 2 \beta\,\sin \alpha\,, \nn \\
g_{\rho^{0}_{1} \rho^{0}_{1} h_{2}} &=& \frac{Y_{\rho\Delta}}{2} \sin2 \beta\,\cos \alpha\,,
\label{dm-h12}
\end{eqnarray}
where $Y_{\rho \Delta}$, $\alpha$ and $\beta$ 
have been defined in the previous section. We had seen in Eq.~(\ref{mix-coup}) that the Yukawa coupling $Y_{\rho \Delta}$ is linearly proportional to $\sin2\beta$ for a given choice of mass splitting $\Delta M_{21}=(M_{\rho^0_2} -  M_{\rho^0_1})$ and vev $v_\Delta$. Therefore, inserting Eqs.~(\ref{mix-coup}) and (\ref{dm-h12}) into Eq.~(\ref{eq:si}) we get 
\begin{eqnarray}
\sigma_{SI} &=& \frac{\mu_{red}^{2}}{\pi}\,\left[ \frac{M_{N} f_{N}}{v} \frac{\Delta M_{21}\,\sin^2 2 \beta\,  \sin 2 \alpha}{4v_\Delta}
\left(\frac{1}{ M_{h_2}^{2}} -
\frac{1}{M_{h_1}^{2}} \right) \right]^{2}
\label{eq:si2}
\end{eqnarray}
Since $\sigma_{SI}$ depends on the model parameters, and since the current limit from DD experiments need to be satisfied, 
they put a constraint on the our model parameter space. Also, the model could be tested and/or  the parameter space can be constrained 
by the future DD experiments like LUX \cite{Akerib:2015rjg, Akerib:2016vxi},
Xenon-1T \cite{Aprile:2015uzo,Aprile:2017iyp}, Panda \cite{Cui:2017nnn} and Darwin \cite{Aalbers:2016jon}.

\subsection{Invisible decay width of Higgs}

If the DM candidate has mass less than half  the
SM-like Higgs mass then the SM-like Higgs could decay to pair of DM particles. This process would contribute to the 
 decay width of the SM-like Higgs into invisible states. The Higgs decay width has been measured very
precisely by the LHC which constrains the 
Higgs decay in such a way that its branching ratio to 
invisible states must be less than 34$\%$ at 95$\%$ C.L.
\cite{Khachatryan:2016vau}.
In the present model the Higgs decay width to invisible states $\rho_1^0$ 
(since in the present work $M_{\rho_{2}^{0}} > \frac{M_{h_1}}{2}$,
hence Higgs can not decay to $\rho_{2}^{0}$) is given by,
\begin{eqnarray}
\Gamma_{h_{1} \rightarrow \rho_{1}^{0}\rho_{1}^{0}} =
\frac{M_{h_1} g_{\rho_{1}^{0}\rho_{1}^{0} h_1}^{2}}{16 \pi} \left(
1 - \frac{4 M_{\rho_{1}^{0}}^{2}}{M_{h_1}^{2}} \right)^{3/2}\,,
\end{eqnarray}
where $g_{\rho_{1}^{0}\rho_{1}^{0} h_1}$
is given in Eq.~(\ref{dm-h12}). 
In order to satisfy the LHC limit, the model parameters have to satisfy the following constraint
\begin{eqnarray}
\frac{\Gamma_{h_{1} \rightarrow \rho_{1}^{0}\rho_{1}^{0}}}{\Gamma_{h_1}^{Total}} \leq 34 \%
{\rm \,\,at \,\,\,95\% \,\,\,C.L.}
\label{eq:higgsdklimit}
\end{eqnarray}
For the parameter range where the kinematical condition $M_{\rho_{1}^{0}} < \frac{M_{h_1}}{2}$ is satisfied we impose the condition given by Eq.~(\ref{eq:higgsdklimit}) and only model parameter values that satisfy this constraints are used in our analysis. 

\subsection{Planck Limit}

Relic density for the DM has been measured very precisely by the satellite
borne experiments WMAP \cite{Hinshaw:2012aka} and Planck \cite{Ade:2015xua}.
In this work we have used the following 
bound on the DM relic density,
\begin{eqnarray}
0.1172 \leq \Omega h^{2} \leq 0.1226 {\rm\,\,\, at \,\,68\% \,\,C.L.}\,,
\end{eqnarray}
which is used to constrain the model parameters such that it is compatible with the Planck limit on DM abundance. 

\section{Dark Matter Relic Abundance  \label{dmresults}}

\begin{figure}[]
	\centering
	\includegraphics[angle=0,height=10cm,width=13cm]{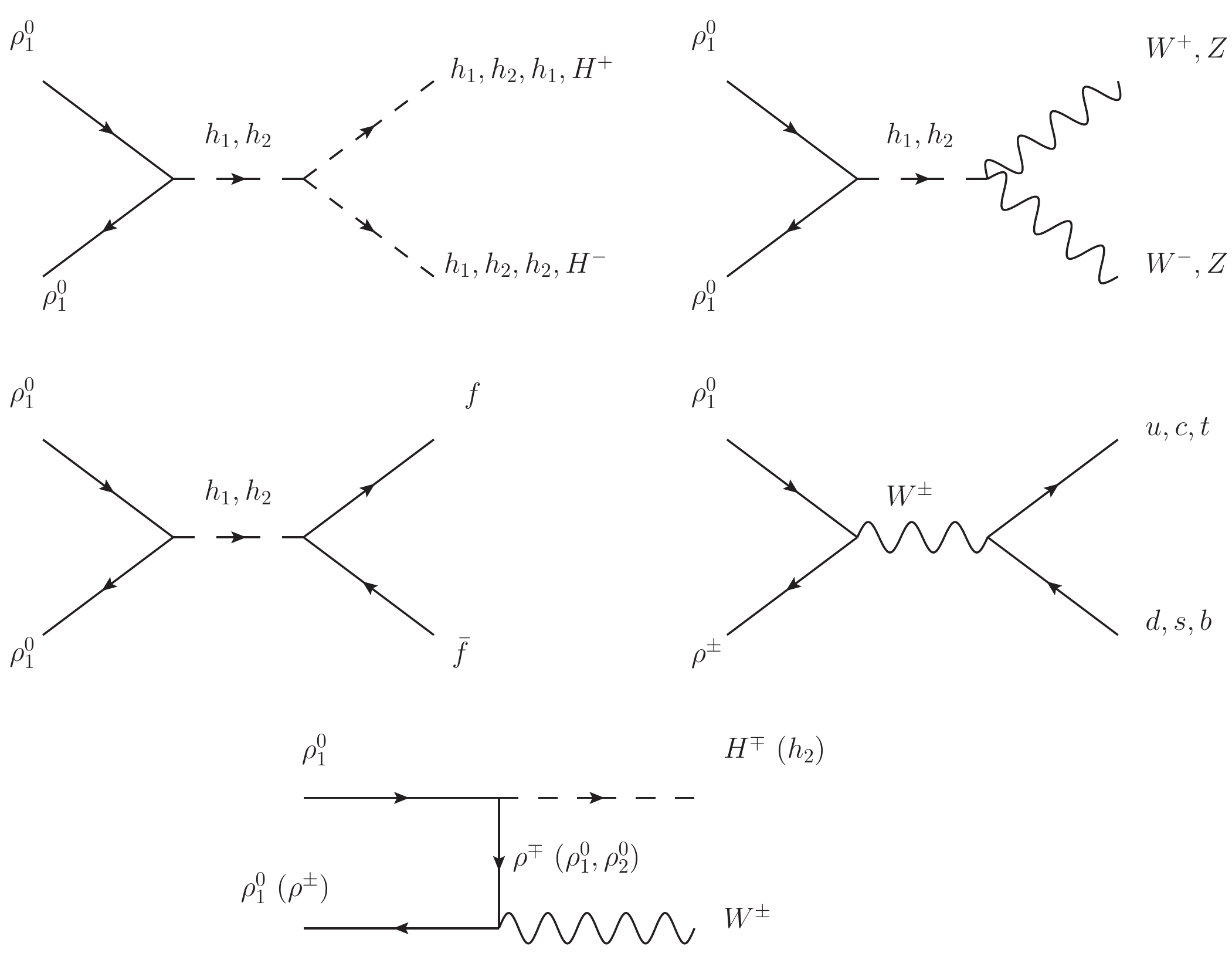}
	\caption{Feynman diagrams which dominantly participate in determining the relic density of DM.}
	\label{feyn-dia}
\end{figure}

\begin{figure}[]
	\centering
	\includegraphics[angle=0,height=7cm,width=8cm]{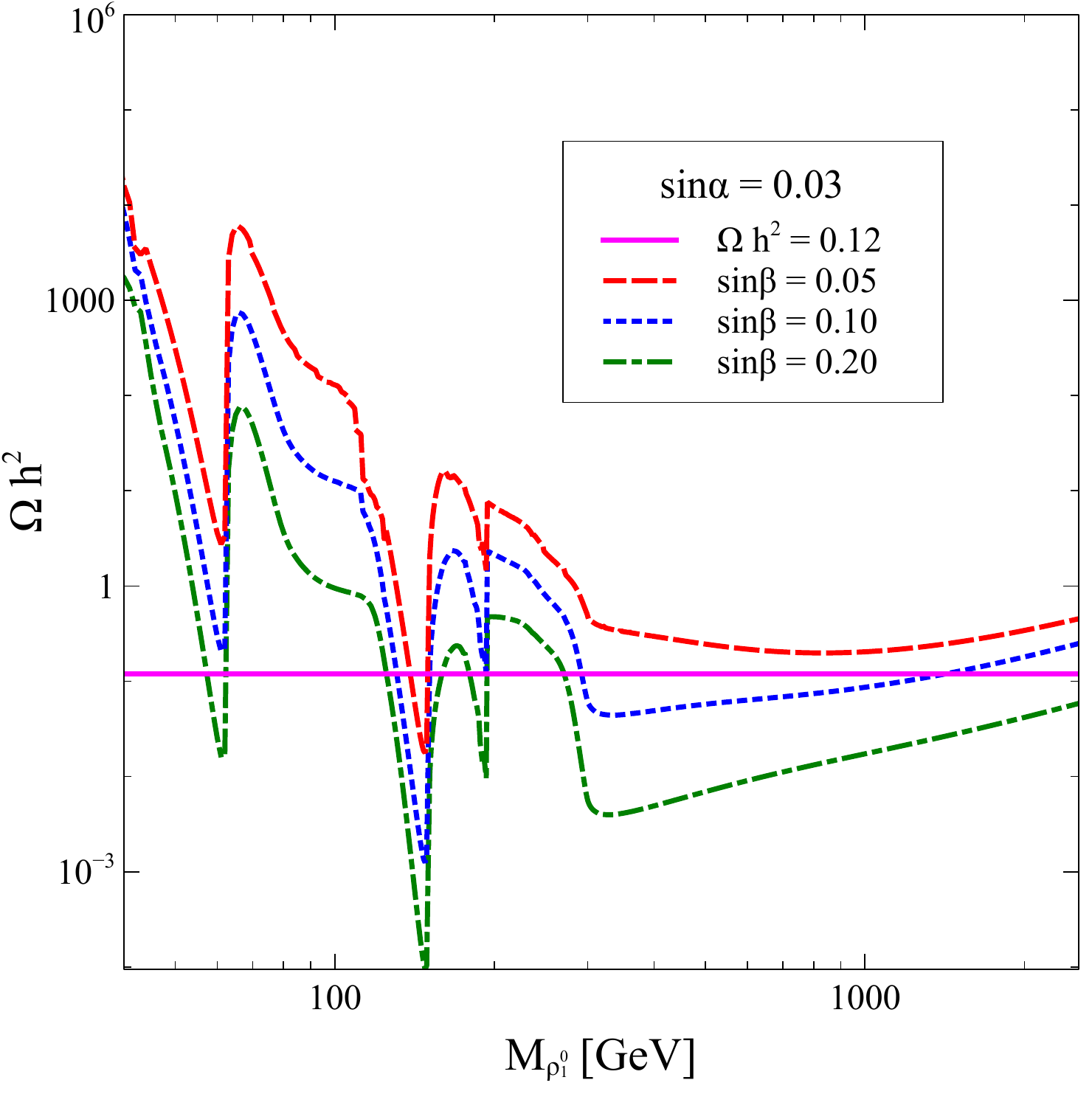}
    \includegraphics[angle=0,height=7cm,width=8cm]{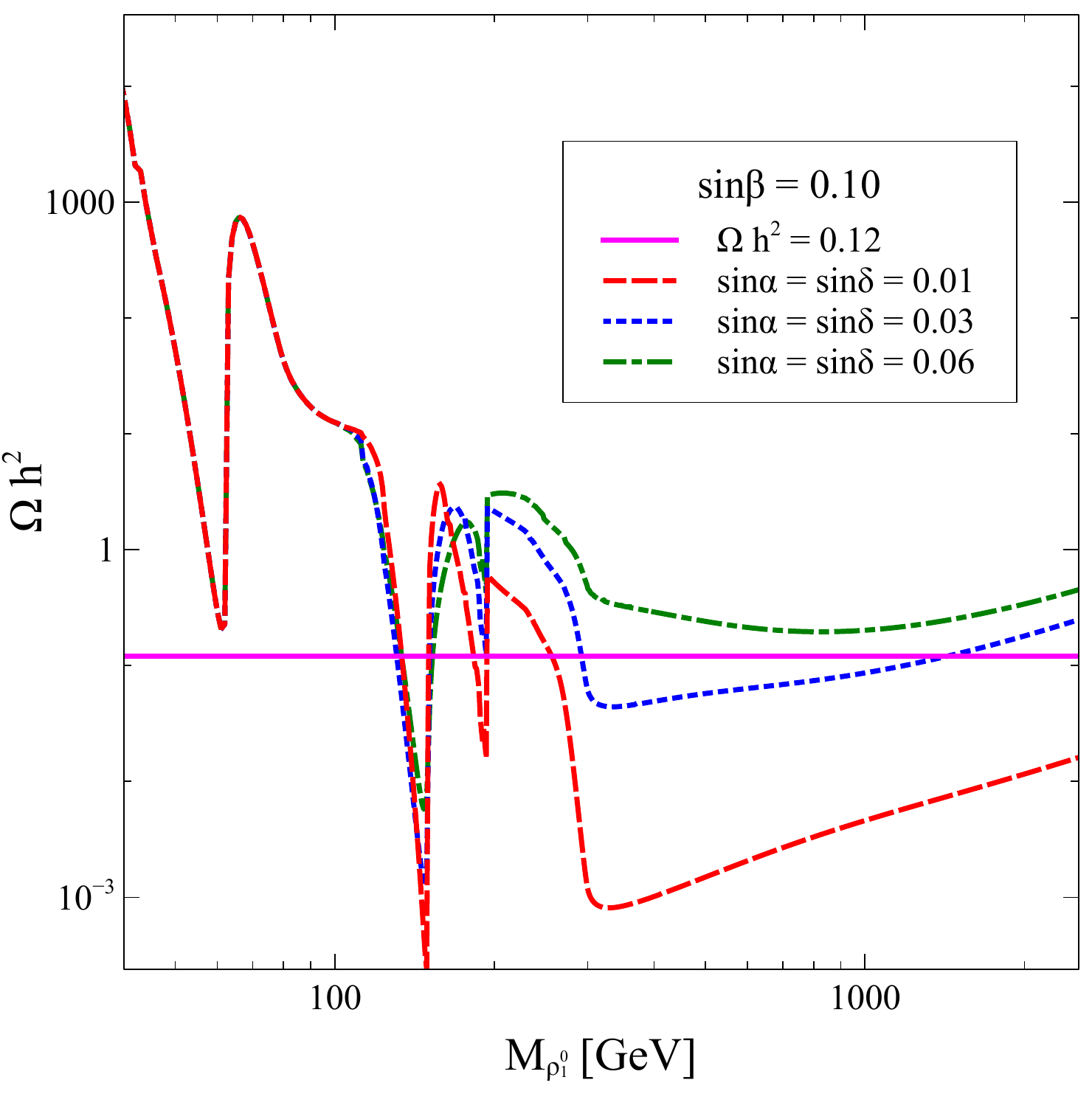}	
	\caption{Left Panel: variation of DM relic density for
	three different values of the singlet triplet fermionic mixing angle
	$\sin\beta$. 
	Right Panel: variation of DM relic density for three different values of the
	neutral Higgses mixing angle $\sin\alpha$. When the BSM
	Higgs value kept fixed at $M_{h_2} = 300$ GeV and we took $\sin\delta$ equal
	neutral Higgs mixing angle for simplicity and kept the mass difference $\Delta M_{12}$
	fixed at 50 GeV.}
	\label{2a}
\end{figure}

In analysing the DM phenomenology we implement the model in
Feynrules  \cite{Alloul:2013bka}. We generate Calchep files using Feynrules
and feed the output files into  micrOmegas \cite{Belanger:2013oya}.
The relevant Feynman diagrams that determine the DM relic abundance are shown in  Fig.\,\ref{feyn-dia}. In presence of triplet as well as singlet states, additional channels mediated by the neutral and charged Higgs state opens up.

\begin{figure}[]
	\centering
	\includegraphics[angle=0,height=7cm,width=8cm]{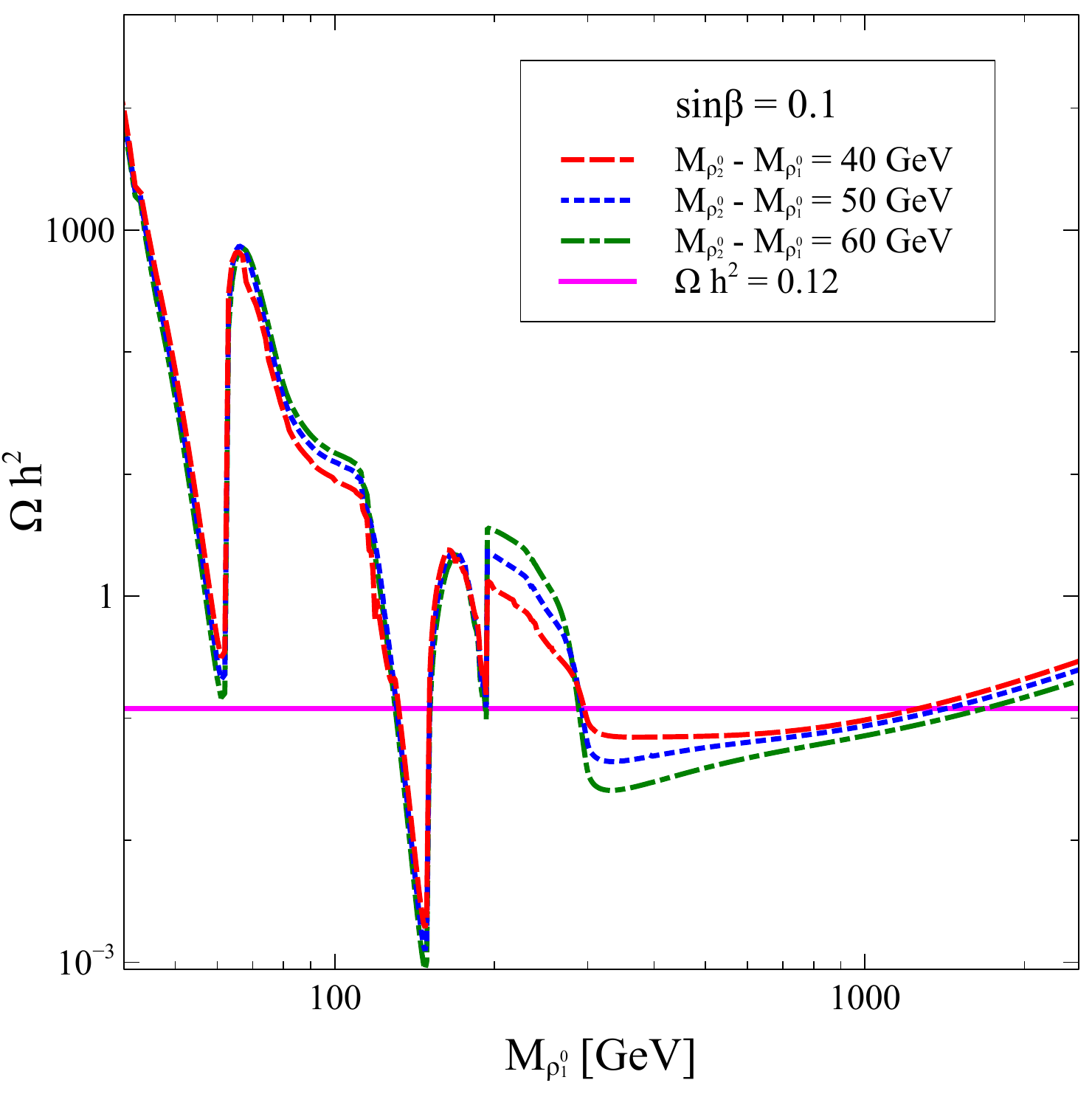}
    \includegraphics[angle=0,height=7cm,width=8cm]{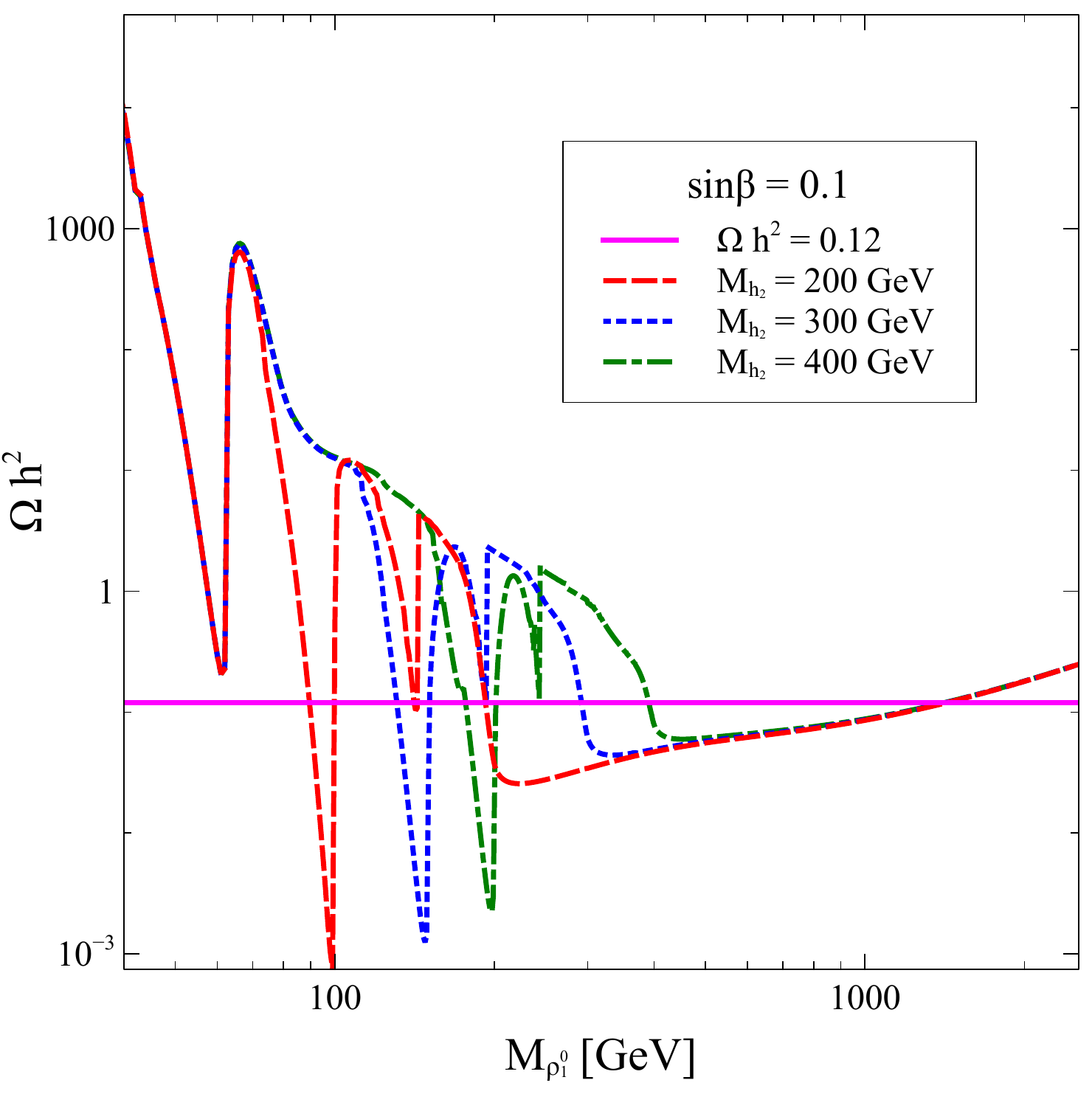}	
	\caption{Left Panel: variation of DM relic density for
	three different value of mass difference $(M_{\rho_{2}^{0}} - M_{\rho_{1}^{0}})$
	when the BSM neutral and charged Higgses values kept fixed at $M_{h_2} = M_{H^{\pm}} =
	300$ GeV. 
	Right Panel: variation of DM relic density for three different value of the
	BSM Higgs mass and we kept the mass difference fixed at
	$M_{\rho_{2}^{0}} - M_{\rho_{1}^{0}} =$ 50 GeV. We took the other parameters value,
	$\sin\alpha = 0.03$, $\sin\delta = 0.03$.}
	\label{3a}
\end{figure}

Different model parameters, such as, the mass of DM, neutral Higgs, mixing  between singlet and triplet fermions, as well as  different Higgs states can impact the 
DM relic density. 
We analyse the dependence of the DM relic density on the model parameters
and also study  the correlation between them that follows from the DM
relic density constraint. Fig.~\ref{2a} shows the variation w.r.t the mass of DM taking into account the variation of the different mixing angles. In other figures, such as, Fig.\,\ref{3a}, we explore the dependency on the mass-difference and the BSM Higgs masses. Few comments are in order:

\begin{itemize}
\item

In the left panel (LP) of Fig.\,\ref{2a}, we show the variation of the 
DM relic density with
 DM mass for three different values of the singlet-triplet mixing angle $\sin\beta$. 
The thin magenta band shows the $2\sigma$ experimentally allowed range of the DM
relic density reported by the Planck collaboration. From the figure, this is evident, that there are four
dip regions with respect to the DM mass. The first resonance  
occurs at $M_{\rho_{1}^{0}} \simeq M_{h_1}/2 \sim 62.5 $ GeV. The SM-like Higgs mediated diagrams shown in  Fig.\,\ref{feyn-dia} give the predominant contribution in this mass range.
The   second resonance occurs at  
$M_{\rho_{1}^{0}} \sim 150$ GeV,  when the DM mass is approximately half the BSM Higgs 
mass ($M_{\rho_{1}^{0}} \simeq M_{h_2}/2$) assumed in this figure. The third dip is due to the t-channel 
diagram $\rho_{1}^{0}\,\rho_{1}^{0} \rightarrow W^{\pm} H^{\mp}$ mediated by the
$\rho^{\pm}$. 
This dip occurs when the DM mass satisfies the relation
$M_{\rho_{1}^{0}} = \frac{M_{W^{\pm}} + M_{H^{\mp}}}{2}$ and happens due to the
destructive interference term of the $W^{\pm} H^{\mp}$ final state. The fourth dip
happens because of the threshold effect of the $W^{\pm} H^{\mp}$ final state
and clear from the fact that with the variation of the
charged scalar mass ($M_{H^{\mp}}$), this dip also changes its position
with respect to the DM mass.
For DM masses greater than this, the DM relic abundance 
is mainly dominated by the s-channel
annihilation diagram where the final state contains $H^{+}H^{-}$, $h_2 h_2$.

\item 
This is to emphasize,  in the present scenario even relatively  lighter DM is in
agreement with 
the observed relic density.  The low mass DM can be copiously produced at LHC and hence can further be tested in the ongoing run of LHC. The lowering of DM mass is possible due to the addition of the 
extra SM gauge singlet fermion $N^{\prime}$ and the extra SM triplet Higgs $\Delta$. This 
opens up  additional annihilation and coannihilation diagrams shown in Fig.\,\ref{feyn-dia}.  
As described before, this  allows  the three resonance regions and make the model compatible 
with the experimental constraint from Planck for DM masses accessible at LHC. This should be contrasted with the 
pure triplet model discussed in section \ref{onlytriplet}, where the 
DM mass compatible with the Planck data is  2.37 TeV, well outside the 
range testable at LHC due to small production cross-section.  
In the next section, we will discuss in detail the prospects of testing the DM at LHC (see Fig.\,\ref{lhc1}). 

\item
The singlet triplet mixing angle $\beta$ has significant effect on the relic density.
 With the increase of the mixing angle $\beta$, the DM relic density
decreases. This happens because the $g_{\rho^0_1 \rho^0_1 h_i}$ ($i=1,2$) coupling increases 
with $\beta$ (cf. Eq.~(\ref{dm-h12})), thereby increasing the cross-section of the annihilation processes. Since the 
relic density is inversely proportional to
the velocity times cross-section $\langle \sigma v \rangle$, where $\sigma$
is the annihilation cross-section of the DM particles and
$v$ is the relative velocity, increase of $\sin\beta$ causes the relic density of DM to decrease.


\item

Additionally, we also explore the effect of the  Higgs mixing angle $\alpha$.
In the right panel (RP) of Fig.\,\ref{2a}, we show the variation of the 
DM relic density for three different values of the doublet-triplet Higgs
mixing angle $\alpha$. The first resonance peak is seen to be nearly unaffected by 
any change in $\sin\alpha$. As the DM mass increases, the impact 
of $\sin\alpha$ increases and we see an increase in the DM relic density with increase 
of $\sin\alpha$. These features can be explained as follows. 
Inserting $Y_{\rho\Delta}$ from Eq.~(\ref{mix-coup}) into Eq.~(\ref{dm-h12}), and replacing 
$v_\Delta$ in terms of $\tan\delta$ using Eq.~(\ref{eq:vdelta}), we get 
\begin{eqnarray}
g_{\rho^{0}_{1} \rho^{0}_{1} h_{1}} &=& \frac{\Delta M_{21} \sin 2 \beta}{2 v}
\frac{\sin\alpha}{\tan \delta}\,, \nn \\
g_{\rho^{0}_{1} \rho^{0}_{1} h_{2}} &=& \frac{\Delta M_{21} \sin 2 \beta}{2 v}
\frac{\cos \alpha}{\tan \delta}\,.
\label{dm-h12_n}
\end{eqnarray}
In our analysis we have taken $\sin\alpha = \sin\delta$ for simplicity. 
Therefore, this results in partial 
cancellations between the the neutral scalars mixing angle and charged scalars mixing
angle, hence we get the following effective couplings for the 
$h_1$ and $h_2$ mediated diagrams, respective,
\begin{eqnarray}
g_{\rho^{0}_{1} \rho^{0}_{1} h_{1}} &=& \frac{\Delta M_{21} \sin 2 \beta}{2 v}
\cos\alpha\,, \nn \\
g_{\rho^{0}_{1} \rho^{0}_{1} h_{2}} &=& \frac{\Delta M_{21} \sin 2 \beta}{2 v}
\frac{\cos^2 \alpha}{\sin \alpha}\,.
\label{dm-h12_n2}
\end{eqnarray}
Since the $h_1$ mediated diagrams effectively depend on $\cos\alpha$ and since 
$\cos\alpha$ remains close to 1 for all the three choices of $\sin\alpha$, taken in 
Fig.\,\ref{2a}, we see no effect of $\sin\alpha$ variation for the $h_1$ 
resonance region. On the other hand, once the $h_2$ mediated diagrams start to dominate, 
the effect of $\sin\alpha$ variation starts to show up. For the $h_2$ resonance region, the 
cross-section decreases as $\sin\alpha$ increases (cf. Eq.~(\ref{dm-h12_n}))
and hence the relic density increases 
with $\sin\alpha$. 
In the vicinity of third resonance region $t$-channel diagrams
dominate and for DM masses
$M_{\rho_{1}^{0}} > M_{h_2},\,M_{H^{\pm}}$,  the $s$-channel
mediated diagrams start contributing in the DM relic density and 
vary the relic density in the expected way with the variation of $\sin\alpha$ and $\sin\delta$. 
\end{itemize}

Additionally, we also show the variation of relic density for different  mass difference $\Delta M_{21}$ in the LP of Fig.\,\ref{3a}. 
The first and second resonance regions show very little dependence on the mass difference $(M_{\rho_{2}^{0}} - M_{\rho_{1}^{0}})$, 
with the relic abundance being marginally less for higher  $(M_{\rho_{2}^{0}} - M_{\rho_{1}^{0}})$. 
However, for the high DM mass we see that the decrease in DM relic abundance with increasing values of 
$(M_{\rho_{2}^{0}} - M_{\rho_{1}^{0}})$ is visible. The reason for this can be understood as follows. 
From  Eq.\,(\ref{mix-coup}), one can
see that the singlet-triplet Yukawa
coupling $Y_{\rho\Delta}$ is directly proportional to the mass difference
$(M_{\rho_{2}^{0}} - M_{\rho_{1}^{0}})$. Both DM couplings
$g_{\rho^{0}_{1} \rho^{0}_{1} h_{2}}$ and $g_{\rho^{0}_{1} \rho^{0}_{1} h_{1}}$ (see Eq.\,(\ref{dm-h12}))
depend on the Yukawa coupling $Y_{\rho\Delta}$ and hence in first and second 
resonance regions where the s-channel processes dominate, 
{\it viz.}, at the resonance regions mainly, controlled by resonance, hence less
effect. On the other hand for higher $M_{\rho^0_1}$ regions and $t$-channel
dominated regions no such resonance exists, so vary linearly with the mass
differences. Close to the 
third resonance region, the t-channel process dominates and here, the cross-section  is suppressed 
due to the propagator mass $M_{\rho^{\pm}}$. Therefore, for regions of the parameter space where the t-channel 
process dominates, the relic abundance is seen to increase as $(M_{\rho_{2}^{0}} - M_{\rho_{1}^{0}})$ (here we considered 
$M_{\rho^{\pm}} - M_{\rho_{2}^{0}} = 160$ MeV) increases for 
a given $M_{\rho^0_1}$. One can see that there is clear cross
over between the $t$-channel and $s$-channel diagrams for
$M_{\rho_{1}^{0}} > M_{H^{\pm}},\,M_{h_2}$, because after this value
of DM mass $\rho_{1}^{0} \rho_{1}^{0}$ mainly annihilates to
$h_{2}h_{2}$ and $H^{+}H^{-}$ by the $s$-channel process mediated by
the Higgses. 

Finally, we also explore the dependency on the mass of the neutral Higgs $h_2$. 
In the RP of Fig.\,\ref{3a},
we show the variation of the relic density with DM mass for three different values of the
BSM Higgs mass: $M_{h_2} = 200$ GeV, 300 GeV and 400 GeV, respectively. From the 
figure we see that the first resonance remains unchanged at 
$M_{\rho_{1}^{0}} \sim 62.5$ because the SM-like Higgs mass is fixed at $M_{h_1} = 125.5$ GeV. 
However, the 
second resonance occurs at three different values of the DM mass depending
on the values of $M_{h_2}$, as the resonance occurs at $M_{\rho_{1}^{0}} \sim \frac{M_{h_2}}{2}$.  
Since here we vary only the BSM Higgs mass $M_{h_2}$,
the couplings which are related to the Higgses remain unaffected, and 
all three curve merge for greater values of DM mass.

To summarise, the relic density depends crucially on the mixing angles between singlet and triplet states, as well as the SM and BSM Higgs, and their masses. The BSM neutral Higgs state with mass 
$M_ {h_2}$ and the charged Higgs state with suitable mass can generate multiple resonance regions, where the DM relic abundance is satisfied. 
The relic abundance varies inversely with the fourth power of $\sin 2 \beta$ {\it i.e.,} $\propto \frac{1}{\sin^{4} 2 \beta}$, where $\beta$ is the singlet-triplet mixing angle.
The DM relic abundance is also seen to depend on the neutral Higgs mixing angle $\alpha$. Below, we discuss the correlation between different model 
parameters.

\begin{figure}[]
	\centering
	\includegraphics[angle=0,height=7cm,width=8cm]{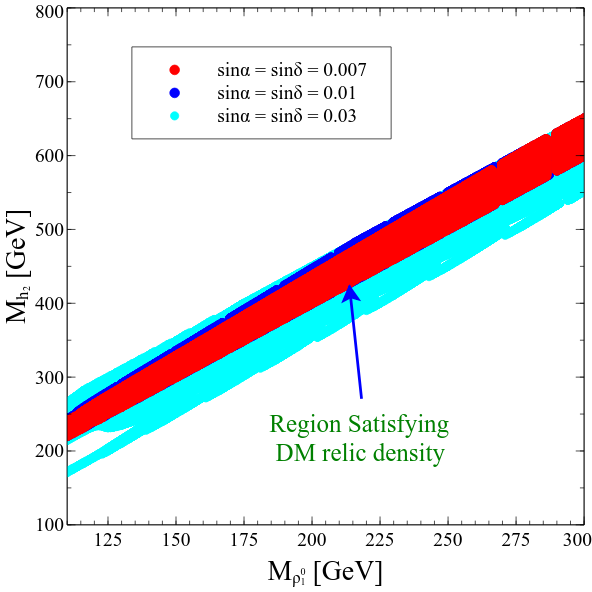}
    \includegraphics[angle=0,height=7cm,width=8cm]{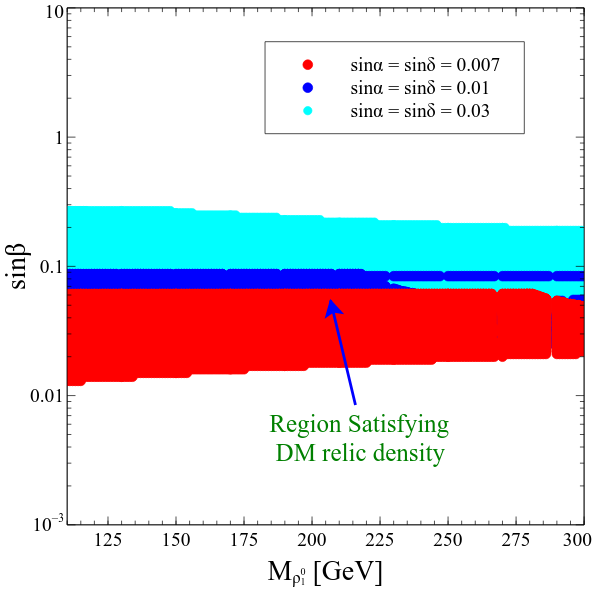}	
	\caption{LP (RP): Allowed region in the $M_{\rho_{1}^{0}}-M_{h_2}$
	($M_{\rho_{1}^{0}} - \sin\beta$) plane after satisfying
	relic density bound. Other parameters values are $\Delta M_{12} = 50$ GeV, $M_{H^{\pm}}
	= M_{h_2}$ and the remaining parameters have been varied as shown in Table \ref{tab-par}.}
	\label{scat1}
\end{figure}
\begin{table}[h!]
\begin{center}
\vskip 0.5cm
\begin{tabular} {||c||c||}
\hline
\hline
Model Parameters & Range\\
\hline
$M_{\rho_{1}^{0}}$ & 110 - 300 [GeV]\\
\hline
$M_{h_2}$ & (2\,$M_{\rho_{1}^{0}}$)$^{100}_{-50}$ [GeV]\\
\hline
$\sin\beta$ &  $10^{-3}$\,\,-\,\,1\\
\hline
\hline
\end{tabular}
\end{center}
\caption{Parametsr varied in the above mentioned range at the time of generating
the scatter plots.}
\label{tab-par}
\end{table} 

\section{Correlation between parameters} 

In the LP of Fig.~\ref{scat1} we show the allowed regions in $M_{h_2}$ and
$M_{\rho_{1}^{0}}$, where all the dots satisfy the relic density bound as
given in Eq.~\ref{dm-bound}. The three colors correspond to three different benchmark choices for the 
Higgs mixing angle $\alpha$. 
From Figs.~\ref{2a} and \ref{3a}, one can
see that the DM relic density can always be satisfied near the resonance regions.
Hence, for a given BSM Higgs mass, there is only a range of DM masses that are allowed 
by the Planck bound. 
In generating the scatter plots we have varied the model parameters
as shown in Table \ref{tab-par}. We have kept the values of $M_{h_2}$
near the resonance region. As expected, we get a sharp correlation
between the mass of DM and the BSM Higgs mass as stressed above. 
On the other hand, in the RP of Fig.~\ref{scat1} we have shown the allowed region
in the sine of singlet-triplet mixing angle ($\sin\beta$) and the
DM mass ($M_{\rho_{1}^{0}}$)  plane. Here we keep $\Delta M_{21} = 50$ GeV
($\Delta M_{21}$ as defined before), and the allowed region shows
that for the given ranges as in Table \ref{tab-par}, the DM relic density can
be satisfied for $0.025 < \sin\beta < 0.27$. 
One interesting point to note here is
that in the LP of Fig.\,\ref{scat1} for $\sin\alpha$, $\sin\delta$ = 0.03,
correlation in the $M_{\rho_1^{0}}-M_{h_2}$
is wider compared to the other two lower values of $\sin\alpha$, $\sin\delta$.
We can understand this as follows. From the RP of Fig.\,\ref{scat1} for $\sin\alpha$,
$\sin\delta$ = 0.03, the DM relic density is satisfied for higher values of $\sin\beta$
($\sim 0.3$).  From the LP of Fig.\,\ref{2a} we see that
near the second resonance region
($M_{\rho_1^{0}} \sim M_{h_2}/2$) the DM relic density is satisfied for a wider range of $M_{\rho^0_1}$ 
for higher values of $\sin\beta$. Since, for $\sin\alpha$, $\sin\delta$ = 0.03,
we get higher values of $\sin\beta$ (as seen from the RP of Fig.\,\ref{scat1}),
so the correlation in $M_{\rho_1^{0}}-M_{h_2}$
planes becomes wider.  

\begin{figure}[]
	\centering
	\includegraphics[angle=0,height=7cm,width=8cm]{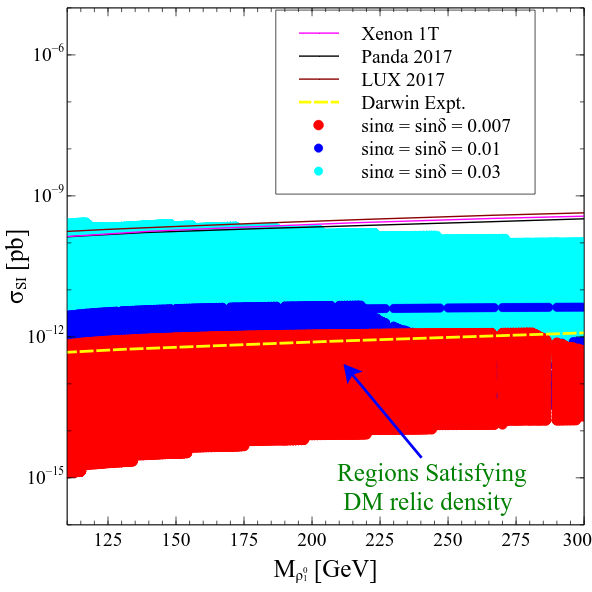}
    \includegraphics[angle=0,height=7cm,width=8cm]{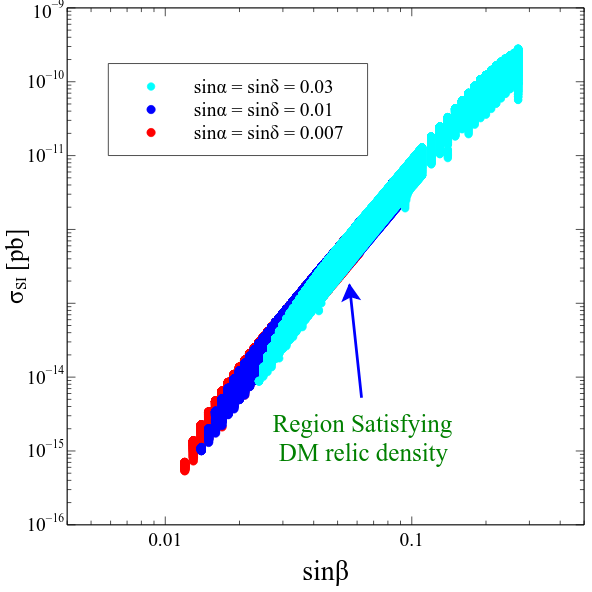}	
	\caption{LP (RP): Allowed region in the $M_{\rho_{1}^{0}}-\sigma_{SI}$
	($\sin\beta-\sigma_{SI}$) plane after satisfying
	relic density bound. Other parameters values are $\Delta M_{12} = 50$ GeV, $M_{H^{\pm}}
	= M_{h_2}$ = 300 GeV and the remaining parameters have been varied as shown in Table \ref{tab-par}.}
	\label{scat2}
\end{figure}

The LP and RP of Fig.\,\ref{scat2} show the allowed regions in the spin independent
DD cross section and the DM mass ($\sigma_{SI} - M_{\rho_{1}^{0}}$) plane 
and the singlet-triplet mixing angle ($\sigma_{SI} - \sin\beta$) plane, respectively.
The LP shows that the model parameter space is not constrained so-far by the results from the 
LUX experiment \cite{Akerib:2016vxi} (and Panda experiment \cite{Cui:2017nnn}).
However, a good part of the parameter space
can be probed in by the Xenon 1T experiment  \cite{Aprile:2017iyp} and in the 
future by the Darwin experiment \cite{Aalbers:2016jon}.
The green, blue and red dots satisfy the present day relic density
bound for three chosen values of $\sin\alpha$. In the RP we show the variation of
the spin independent direct detection
cross-section 
with the fermion singlet-triplet mixing angle. Since the DD cross section is directly proportional to the
square of $\sin\beta$ we see this functional dependence in this figure and $\sigma_{SI}$ is seen to increase with 
$\sin\beta$. 

\section{Indirect Detection of Dark Matter by $\gamma \gamma$ observation}
\label{indirect-detection}
\begin{figure}[]
	\centering
	\includegraphics[angle=0,height=9cm,width=18cm]{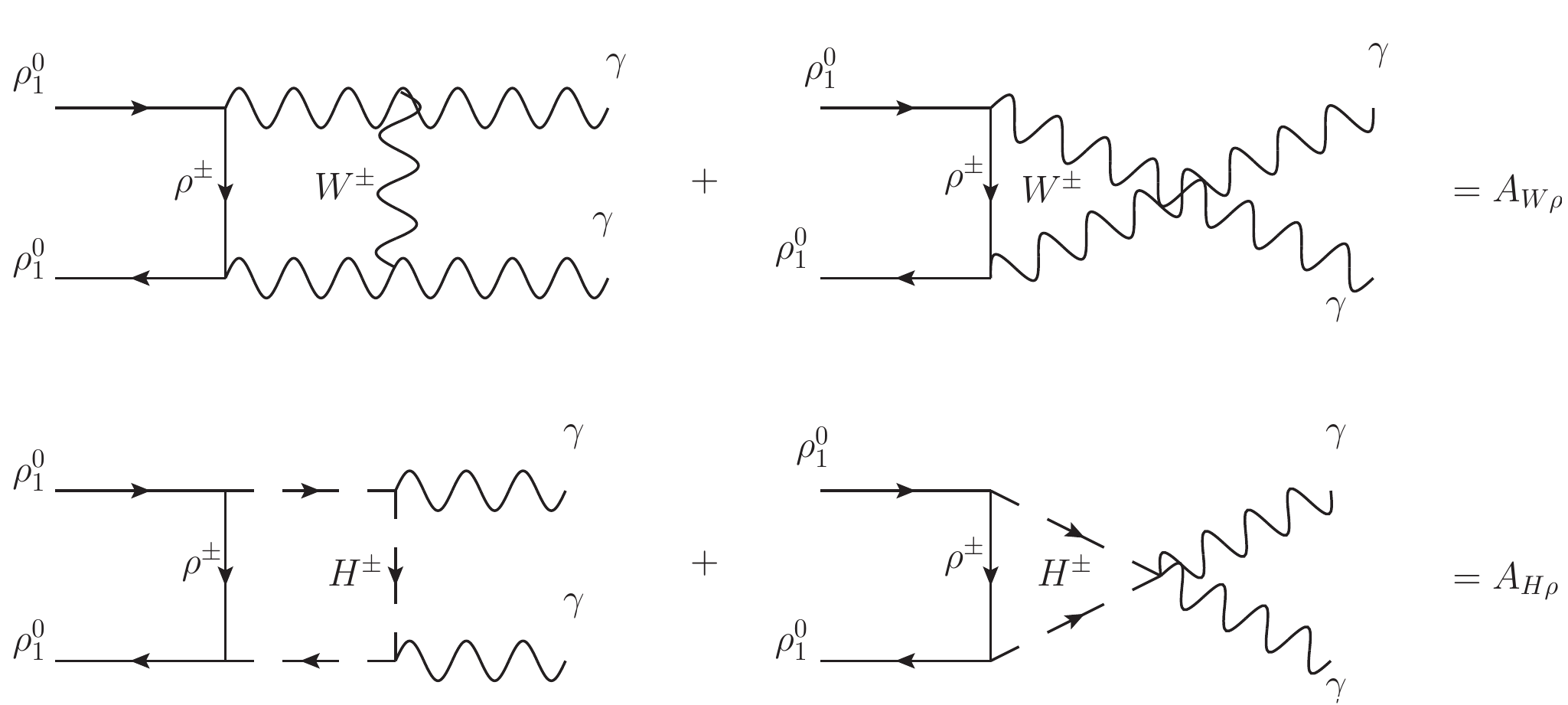}
	\caption{Feynman diagrams of the DM annihilation into the gamma rays by one loop
	diagrams mediated by the charge gauge boson $W^{\pm}$ and the charged scalar
	$H^{\pm}$.}
	\label{feyn-dia-gamma-rays}
\end{figure}
In addition to the detection of the DM in the ongoing direct detection experiments
for the present model, it can also be detected by the indirect search of DM in
different satelite borne experiments
like Fermi-LAT \cite{Ackermann:2013uma, Ackermann:2015lka},
HESS \cite{Abramowski:2013ax, Profumo:2016idl}
by detecting the gamma-rays signal which comes from the
DM annihilation. In the present
situation DM cannot annihilate to gamma-rays at tree level but certainly
can annihiate at the one loop level mediated by the charged gauge boson $W^{\pm}$
and the charged scalar $H^{\pm}$ which is shown in
Fig.\,\ref{feyn-dia-gamma-rays}. The average of the amplitude for
the velocity times cross section for the Feynman diagrams which are shown in
Fig.\,\ref{feyn-dia-gamma-rays}
takes the following form \cite{Bergstrom:1997fh, Bern:1997ng}, 
\begin{eqnarray}
\langle \sigma v \rangle_{\gamma \gamma} = \frac{\alpha_{EM}^2 M^2_{\rho_1^0}}{16 \pi^3}
|A|^2
\end{eqnarray}
where $A = A_{W\rho} + A_{H\rho}$, $\alpha_{EM} = e^2/4\pi$ and $e = 0.312$.
$A_{W\rho}$ and the $A_{H\rho}$ are the separate contribution from one loop
diagrams as shown in Fig.\,\ref{feyn-dia-gamma-rays} which are mediated by
the $W^{\pm}$ and $H^{\pm}$, respectively. 
The individual amplitude for the diagrams which are shown in
Fig.\,\ref{feyn-dia-gamma-rays} take the following form,
\begin{eqnarray}
A_{W\rho} &=& -2 C_1^2 \left[ 2 I_3^a(M_{W}) +
2(M_{\rho^{\pm}}^2 + M_{W}^2 - M_{\rho_1^0}^2)I_4^a + 2 M_{\rho^\pm}^2 I_4^b
+ 3 M_{\rho^{\pm}}^2 I_4^c + I_3^b(M_{W},M_{\rho^{\pm}}) \right] \nn \\
&& + 8 C_1^2 M_{\rho^{\pm}} M_{\rho_1^0}(I_4^b + I_4^c), \nn \\
A_{H\rho} &=& C_2^2 \left[ 2 M_{\rho^{\pm}}^2 I_4^b + M_{\rho^{\pm}}^2 I_4^c
+ I_3^b (M_{H^{\pm}}, M_{\rho_1^0}) \right] 
\label{amplitude-gamma-rays}
\end{eqnarray}
where the explicit form of $I_3^i$ ($i=a,b$) and $I_4^j$ ($j = a,b,c$)
are given in the Appendix. The couplings $C_1$ and $C_2$ are $C_1 = -e \sin \beta/\sin \theta_w$, where
$\theta_w$ is the Weinberg angle and $C_2 = \cos\alpha Y_{\rho \Delta}/2$,
where $Y_{\rho\Delta}$ is given in Eq.\,(\ref{mix-coup}).

\begin{figure}[]
	\centering
	\includegraphics[angle=0,height=9cm,width=12cm]{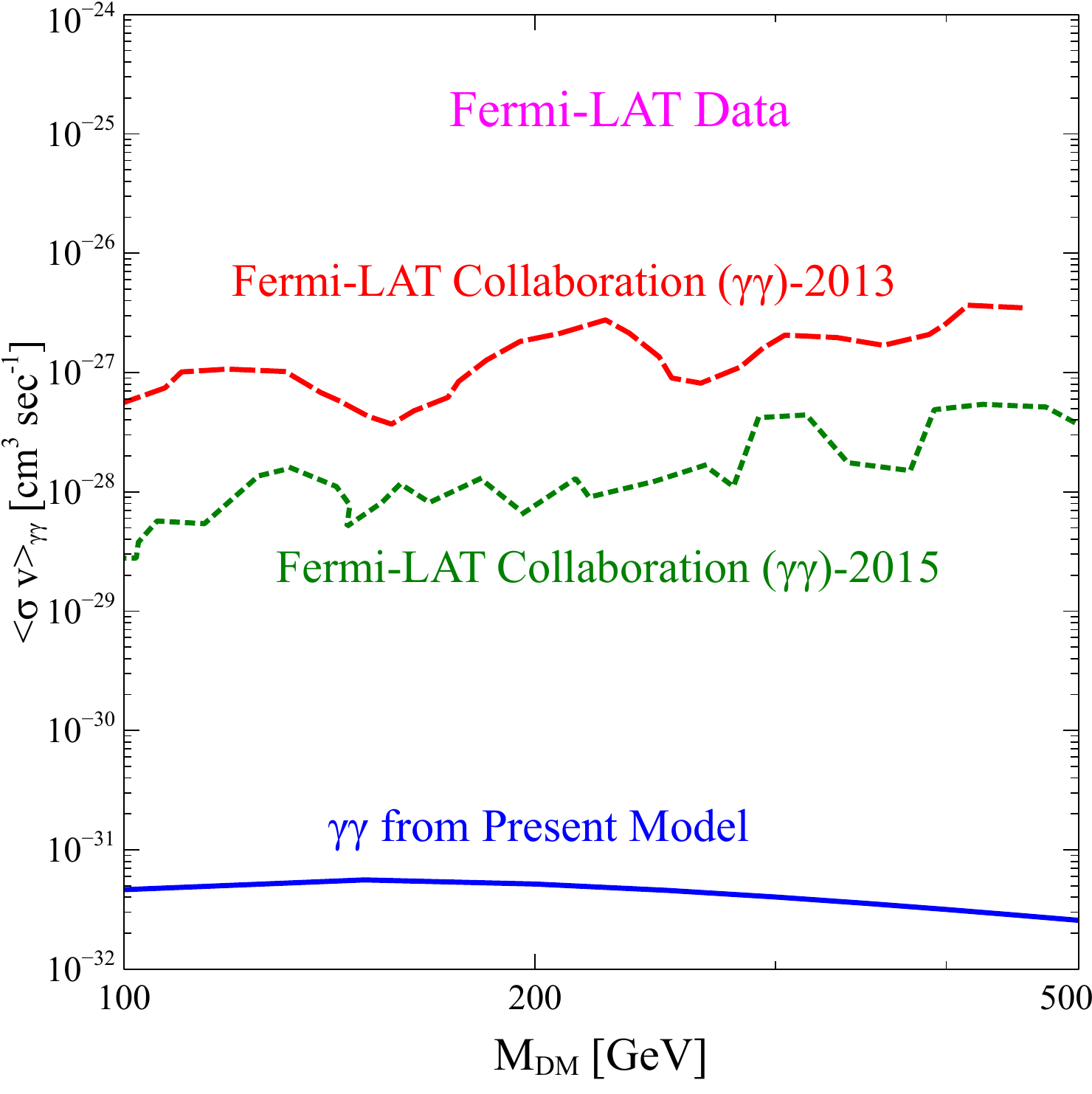}
	\caption{Fermi-LAT bounds and the prediction from the present model. In
	getting the prediction from the model,
	we have kept the parameters value fixed at $\sin \beta = 0.1$, $\Delta M_{21} = 50$ 
	GeV, $M_{H^{\pm}} =  M_{h_2} = 2M_{DM} $  and $\sin \alpha = \sin \delta = 0.03$.}
	\label{gamma-rays}
\end{figure}

In Fig.\,\ref{gamma-rays}, we show the variation of $\langle \sigma v \rangle$
with the DM mass, $M_{\rho_1^0}$ by considering the relevant one loop
diagrams. As the DM relic density for the pure triplet fermion is 
satisfied for DM mass of about $2.4$ TeV and this is already ruled out by the
Fermi-LAT data when we the Sommerfeld enhancement  is taken into consideration.
In the current work, we have taken the triplet fermion mixing with
the singlet fermion with the help of the triplet scalar and
DM relic density can be satisfied around the
100 GeV order DM mass. For such low mass range of the DM where
$M_{\rho_1^0} \sim M_{W} \sim M_{H^{\pm}}$, the Sommerfeld enhancement factor
will have no significant role in the increment
of $\langle \sigma v \rangle_{\gamma \gamma}$. We have shown the Fermi-LAT-2013
\cite{Ackermann:2013uma}
and Fermi-LAT-2015 \cite{Ackermann:2015lka} data in the
$\langle \sigma v \rangle_{\gamma \gamma} - M_{\rho_1^0}$ plane by the red and green
dash line, respectively. By blue solid line we have shown the $\langle \sigma v \rangle_{\gamma \gamma}$ variation with the DM mass which is suppressed
by the one loop factor for the present model.


\section{LHC Phenomenology}
\label{lhc}
Although there has been no dedicated search for such a model at the LHC, one can in principle, derive 
limits on the masses of the exotic fermions ($\rho^0_{1,2}$, $\rho^{\pm}$) and the additional scalar 
states($h_2$, $H^{\pm}$) from existing LHC analyses looking for similar particles. 
LHC has extensively searched for heavy neutral Higgs boson similar to $h_2$ and the non-observation 
of any such states puts stringent constraints on masses and branching ratios of such particles provided 
their decay modes are similar to that of the SM-like Higgs \cite{Aaboud:2017gsl,Aaboud:2017rel,Aaboud:2017yyg}. 
However, in our case, these bounds are significantly weakened because the decays of $h_2$ here are quite 
different compared to the conventional modes. $h_2$ mostly decays into $h_1h_1$, $\rho^0_2\rho^0_1$ or $\rho^0_1\rho^0_1$
pair depending on the availability of the phase space. In absence of $\rho^0_2\rho^0_1$ mode, $\rho^0_1\rho^0_1$ always 
has a large (30\% - 40\%) branching ratio, which is a completely invisible mode and thus leads to weaker event rates in the visible final 
states. In the presence of $\rho^0_2\rho^0_1$ and(or) $h_1h_1$ modes, a $b\bar b$ final state study can constrain the $h_2$ mass since $\rho^0_2$ 
always decays dominantly via $b\bar b$. However, the net branching ratio suppression results in weaker limits from the existing studies.   
Charged Higgs search at the LHC concentrates on the $\tau\bar\nu$, $c\bar s$, $c\bar b$ and $t\bar b$ decay modes depending on the mass of 
$H^{\pm}$ \cite{Khachatryan:2015qxa,Khachatryan:2015uua,CMS:2016qoa,ATLAS:2016qiq}. None of these decay modes are significant 
in our present scenario. Here $\rho^{\pm}$ decays via $\rho^0_1\rho^{\pm}$ and(or) $W^{\pm}Z$ depending on the particle masses.
Thus the existing charged Higgs mass limits do not apply here. Instead, a dilepton or trilepton search would be more suitable 
for such particles although the charged leptons originating solely from the gauge boson decays will be hard to distinguish 
from those coming from the SM. 
Constraints on the masses of $\rho^0_2$ and $\rho^{\pm}$ can be drawn from searches of wino-like neutralino 
and chargino in the context of supersymmetry \cite{ATLAS:2017uun,Aaboud:2017nhr}. However, production cross-section 
of this pair at the LHC is smaller compared to the gauginos leading to weaker mass limits. Moreover, the decay 
pattern of $\rho^0_2$ is quite different from that of a wino-like neutralino. The most stringent gaugino mass 
bounds are derived from the trilepton final state analysis. Such a final state cannot be expected in our present scenario
since $\rho^0_2$ dominantly decays into a $b\bar b$ pair along with $\rho^0_1$. However, $\rho^{\pm}$ always decays into $\rho^0_1$ 
associated with an on-shell or off-shell $W$-boson, similar to a wino-like chargino. Thus the bounds derived on the chargino masses 
in such cases \cite{ATLAS:2017uun,Aaboud:2017nhr} can be applied to $m_{\rho^{\pm}}$ as well if appropriately scaled to its 
production cross-section and subjected to $m_{\rho^0_1}$. We have taken this constraint into account while constructing our benchmark points.

In this section, we discuss in detail the LHC phenomenology of the dark matter. The low mass dark matter can be copiously produced at LHC, either directly or from the decay of the its triplet partner.  
\subsection{Production cross-section and choice of benchmark points}
\label{sec:bp_prod}
\begin{figure}[h!]
	\centering
	\includegraphics[angle=0,height=8cm,width=8cm]{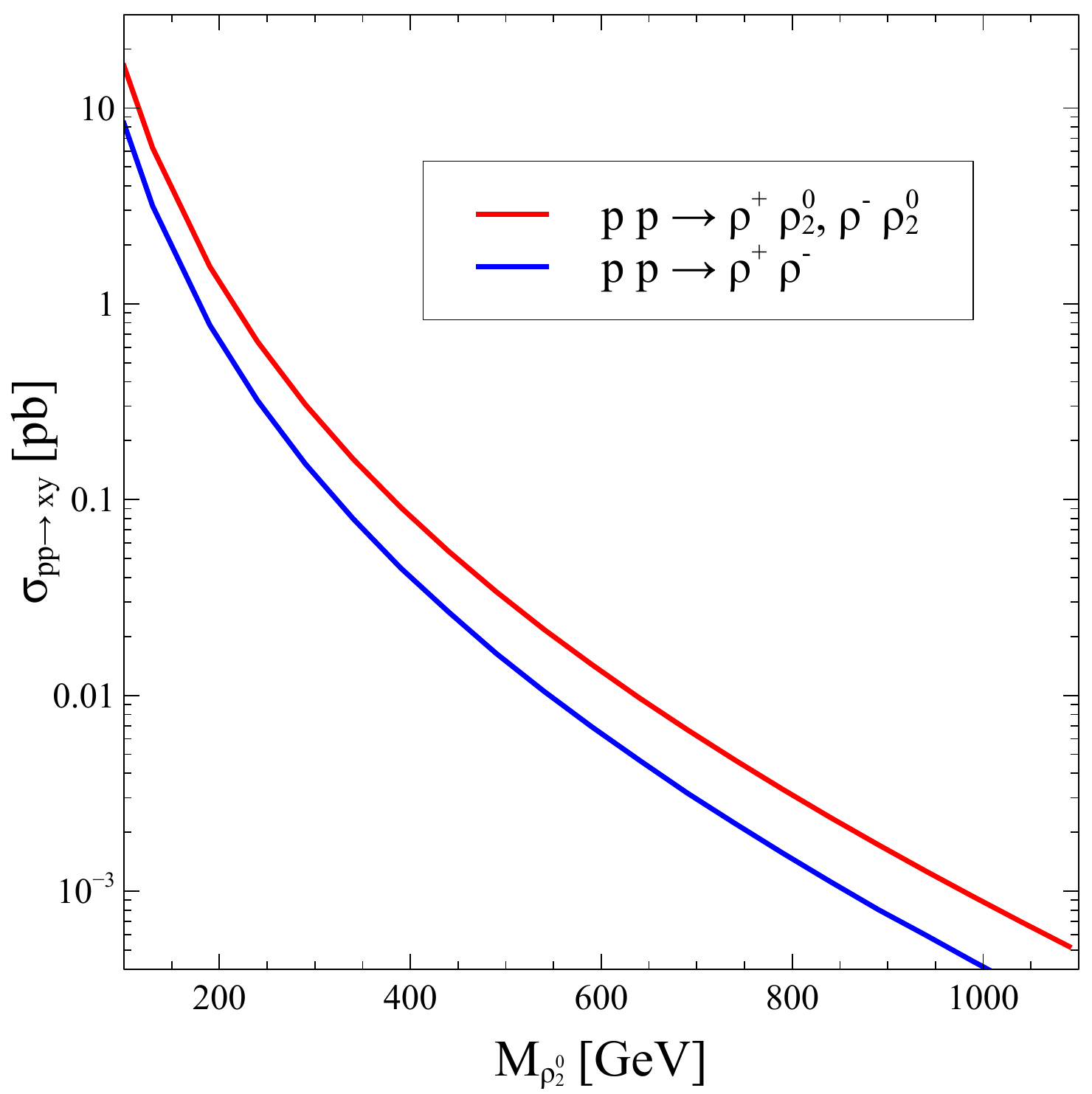}
	\caption{Variation of production cross section $\rho^{\pm} \rho_{2}^{0}$ and
	 $\rho^{\pm} \rho^{\mp}$ with DM mass for
	13 TeV run of LHC where we kept fixed $M_{\rho_{2}^{0}} - M_{\rho_{1}^{0}} = 20$ GeV,
	$M_{h_2} = M_{H^{\pm}} = 300$ GeV.}
	\label{lhc1}
\end{figure}

For this we consider production of $\rho^{\pm} \rho_{2}^{0}$ which further decay into $\rho_{1}^{0}$ associated with 
quarks resulting in a multi-jet + $\cancel{E}_{T}$ signal. Similar collider signal can also arise from other production 
modes, namely, $\rho_{2}^0 \rho_2^0$ and $\rho^+\rho^-$. While the $\rho_2^0$ pair production cross-section is smaller 
by orders of magnitude, the other two production channels have comparable cross-sections as shown in Fig.~\ref{lhc1}. 
For Fig.~\ref{lhc1}, we have kept the mass gap between
$\rho_2^{0}$ and $\rho_{\pm}$ fixed at the pion mass and the cross-section is computed 
at 13 TeV centre-of-mass energy.
Clearly, $\sigma (pp\to \rho^{\pm} \rho_{2}^{0})$ is almost twice to that of $\sigma (pp\to \rho^+\rho^-)$ making the 
former one the most favored production channel to probe for the present scenario. However, the latter one can also 
contribute significantly to boost the multi-jet + $\cancel{E}_{T}$ signal event rate given the fact that $\rho_2^0$ 
and $\rho^{\pm}$ are mass degenerate from the collider perspective. The degeneracy of $\rho_2^0$ 
and $\rho^{\pm}$ results in their decay products to have very similar 
kinematics. Therefore, in our study of the multi-jet final state we have included both the production channels $pp\to\rho_2^0\rho^{\pm}$ and 
$pp\to\rho^+\rho^-$.  $\rho_2^0$ further decays into $\rho_1^0$ mostly via $h_2$ whereas 
$\rho^{\pm}$ also decays into $\rho_1^0$ via $W$-boson. Regardless of whether the intermediate scalar or the gauge bosons are 
on-shell or off-shell, we always consider their decays into pair of $b$-quarks or light quarks. In the former case, the decay 
of $h_2$ is likely to give rise to $b$-jets in the final state whereas the latter one results in light jets arising from $W$ decay.  
Hence in order to combine the event rates arising from these two production channels, we do not demand any 
$b$-tagged jets in the final states. Besides, demanding $b$-tagged jets in the final state can also hinder the signal event rates 
specially for cases where the mass difference between $\rho^0_2$ and $\rho^0_1$, i.e., $\Delta M_{21}$ is small. 

For detail collider simulation and analysis of the above mentioned final state,
we have constructed few benchmark points representative of the available parameter space after imposing all the 
relevant constraints. We have presented 
our choice of benchmark points with all the relevant particle masses and DM constraints in Table~\ref{tab3}. 
\begin{table}[h!]
\begin{center}
\vskip 0.5cm
\begin{tabular} {||c|c|c|c|c|c|c|c||}
\hline
\hline
Parameters & $M_{\rho^{0}_{1}}$ [GeV] & $M_{\rho^{0}_{2}}$ [GeV] & $M_{\rho^{+}}$ [GeV] &
$M_{h_2}$ [GeV] & $M_{H^{\pm}}$ [GeV] & ~~~$\sigma_{SI}$ [pb]~~~
&~~~ $\Omega h^{2}$~~~\\
\hline
\hline
BP1 & 87.6 & 128.0 & 128.2 & 195.5 & 195.5 & 2.1 $\times 10^{-12}$ & 0.1207 \\
\hline
BP2 & 132.0 & 172.0 & 172.2 & 300.0 & 300.0 & 4.1 $\times 10^{-12}$  & 0.1208\\
\hline
BP3 & 171.1 & 211.0 & 211.2 & 400.0 & 400.0 & 4.8 $\times 10^{-12}$ & 0.1197\\
\hline
BP4 & 86.7 & 200.0 & 200.2 & 194.1 & 194.1 & 1.8 $\times 10^{-11}$ & 0.1186\\
\hline
BP5 & 119.0 & 230.0 & 230.2 & 280.0 & 280.0 & 2.9 $\times 10^{-11}$ & 0.1195\\
\hline
\hline
\end{tabular}
\end{center}
\caption{Benchmark points to study LHC phenomenology. We fixed other BSM parameters
as $\sin\alpha = 0.03$, $\sin \beta = 0.1$.}
\label{tab3}
\end{table}
Note that, the mass gap between $\rho_2^0$ (or $\rho^{}\pm$) 
and $\rho_1^0$ ($\Delta M_{21}$) can not be arbitrarily large for admissible values of $\beta$. Hence in some cases, these fermionic 
states can lie quite close together giving rise to a {\it compressed} scenario as depicted by, for example, BP1 in 
Table~\ref{tab3}. However, the mass gap can be moderate to significantly large and our choice of the benchmark points 
encompasses all possible kind of DM mass regions and mass hierarchies. 
\subsection{Simulation details }
\label{sec:lhc_simul}

As mentioned previously, the 
mass gap $\Delta M_{21}$ can be quite small in some cases, resulting in soft jets in the final state, which may escape detection. 
The standard procedure is to tag the radiation jets in order to look for such scenarios. For that, one needs to take into account 
production of the mother particles along with additional jets and perform a proper jet-parton matching \cite{Mangano:2006rw, Hoche:2006ph}
in order to avoid double counting 
of jets. We have considered the above mentioned production channels associated with upto two additional jets at the parton level. 
\begin{eqnarray}
p\, p &\rightarrow& X Y\nn \\
p\, p &\rightarrow& X Y j\nn \\
p\, p &\rightarrow& X Y j j 
\end{eqnarray}
where \{X Y\} indicates any of the three pairs, \{$\rho_2^0$ $\rho^+$\}, \{$\rho_2^0$ $\rho^-$\} and \{$\rho^+$ $\rho^-$\}.
The events have been generated at the parton level using MadGraph5(v2.4.3) \cite{Alwall:2014hca, Alwall:2011uj} with 
CTEQ6L \cite{Pumplin:2002vw} parton distribution function (PDF). 
Events were then passed through PYTHIA(v6.4) \cite{Sjostrand:2006za} to perform
showering and hadronisation  effects.  Matching between the shower jets and the parton level jets has been done using
MLM \cite{Mangano:2006rw, Hoche:2006ph} matching scheme. We have subsequently passed the events through
Delphes(v3.4.1) \cite{deFavereau:2013fsa, Selvaggi:2014mya, Mertens:2015kba} for jet formation based on the anti-$k_{T}$ jet clustering
algorithm \cite{Cacciari:2008gp} via fastjet \cite{Cacciari:2011ma} and for detector
simulation we used the default CMS detector cuts.

Since the number of hard jets obtained in the cascade are expected to vary for the different benchmark points depending on the 
choice of $\Delta M_{21}$, we have chosen our final state with an optimal number of jet requirement along with missing energy: 
$\ge 2$-jets + ${\cancel E_T}$. 
The dominant SM background contributions for such a signal can arise from $QCD$, $V$+ jets, $t\bar t$+ jets and 
$VV$ + jets channels, where $V = W^{\pm}$ and $Z$.
For collider analysis of this final state we have followed strategy similar to that adopted in, for example \cite{Aad:2014wea, Dutta:2015exw}.  
\subsubsection*{\bf Selection Cuts}
\label{sel_cuts}
We have used the following basic selection cuts (A0) to identify the charged leptons (e, $\mu$),
photon ($\gamma$) and jets in the final state: 
\begin{itemize}
\item Leptons are selected with $p_{T}^{l} > 10$ GeV and the pseudorapidity
$|\eta^{\ell}| < 2.5$, where $\ell = e, \mu$.
\item We used $p_{T}^{\gamma} > 10$ GeV and psudorapidity $|\eta^{\gamma}| < 2.5$
as the basic cuts for photon.
\item We have chosen the jets which satisfy $p_{T}^{j} > 40$ GeV and $|\eta^{j}| < 2.5$.
\item We have considered the azimuthal separation between  all reconstructed
jets and missing energy must be greater than 0.2 i.e. $\Delta \phi(jet,\cancel{\vec E}_{T}) >
0.2$.
\end{itemize}

\begin{figure}[]
\centering
\includegraphics[angle=0,height=7cm,width=8cm]{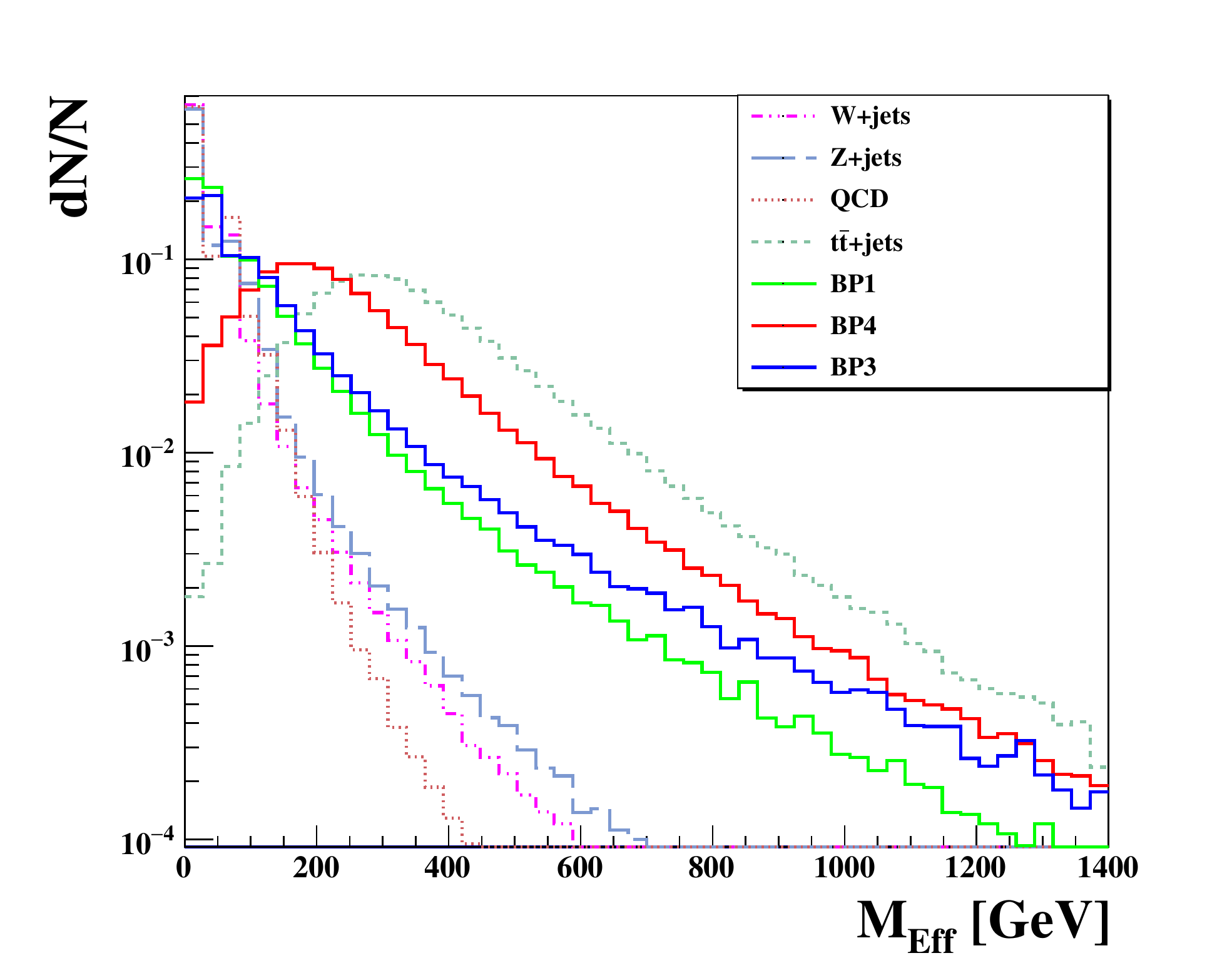}
\includegraphics[angle=0,height=7cm,width=8cm]{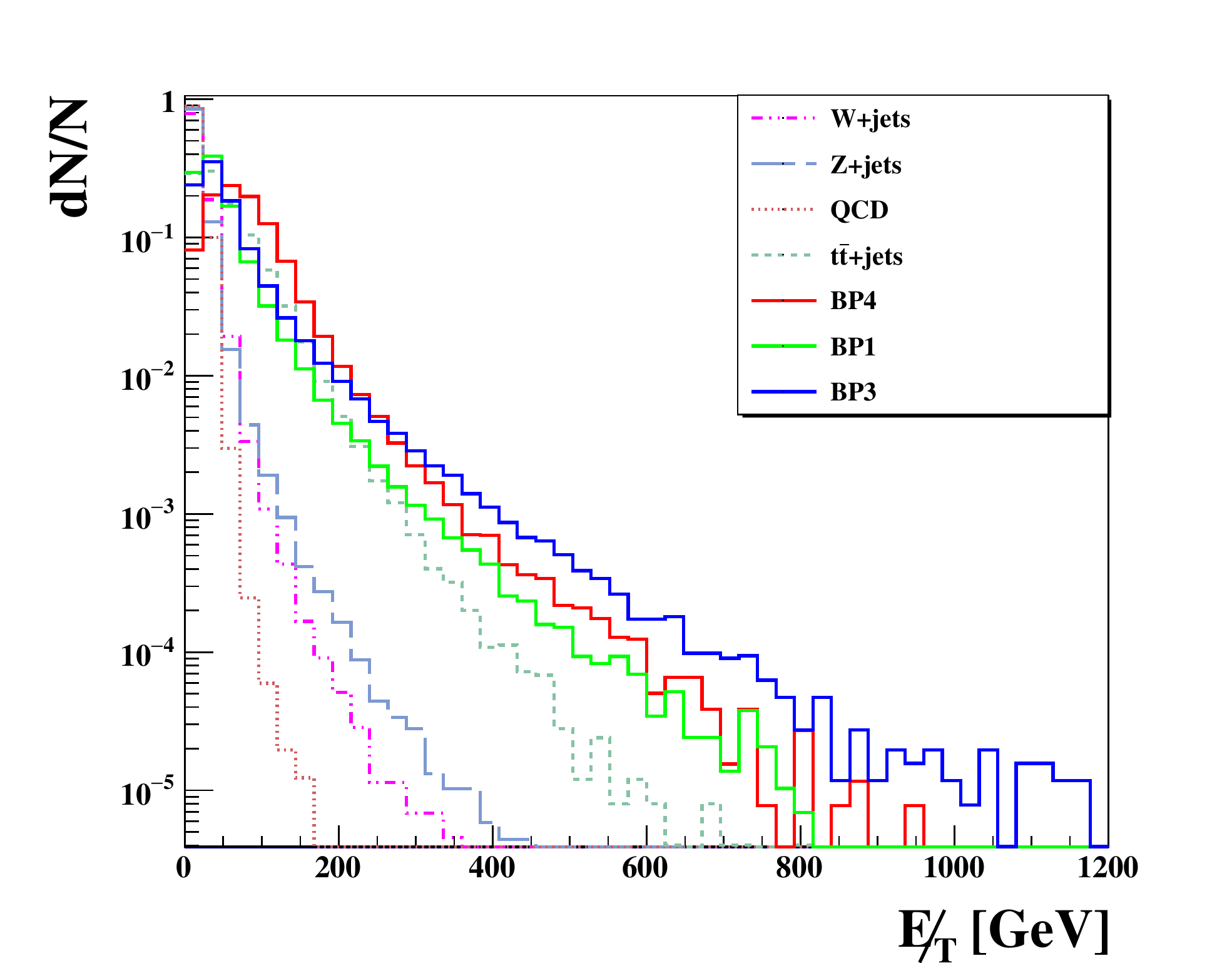}	
\caption{Normalised differential distribution with respect
to the different cuts which we have used in our study. Besides the SM backgrounds
we have also shown the distribution of three benchmark points BP1, BP4, BP3.
All the kinematic variables have been addressed in text.}
\label{histo}
\end{figure}

In Fig.\,\ref{histo}, we have shown the distribution function of different kinematic variable 
for the illustrative benchmark points 
 after applying the basic selection cuts (A0). In addition, we also show the distribution for the SM background events.
The signal event distributions shown here correspond to $\rho_2^0\rho^{\pm}$ 
production channel which is the dominant contributor to the final state. 

Here we have shown the distribution corresponding to the effective mass ($M_{Eff}$) and
missing energy ($\cancel{E_{T}}$) where the effective mass defined in the following way,
\begin{eqnarray}
M_{Eff} = \sum_{i} |\vec{p}_{T_i}^{j}| + \sum_i |\vec{p}_{T_i}^{\ell}| + \cancel{E_{T}}
\end{eqnarray} 
These distributionis show some distinguishing features of the signal events from the SM backgrounds. 
Guided by these distributions we now proceed to device some appropriate kinematic cuts 
to optimise the signal to background events ratio in order to maximise the statistical significance of the 
signal.
\begin{enumerate}
\item[A1:] Since we are studying a hadronic final state, we have imposed a lepton and
photon veto in the final state. This cut coupled with a large ${\cancel E_T}$ cut helps 
to reduce background events particularly arising from $W$ + jets when $W$ decays leptonically. 
\item[A2:] $p_T$ requirements on the hardest and second hardest jets: 
$p_{T}^{j_1} > 130$ GeV and $p_{T}^{j_2} > 80$ GeV. This cut significantly reduce the
V + jets (where V = $W^{\pm}$, Z) and QCD backgrounds.
\item[A3:] The $QCD$ multi-jet events have no direct source of missing energy. Therefore, 
any contribution to ${\cancel E_T}$ in these events must arise from the mismeasurement of the 
jet $p_T$s. In order to minimise this effect, we have ensured that the $\cancel{\vec E}_T$ and the 
jets are well separated, i.e., $\Delta \phi(j_i,\cancel{\vec E}_T) > 0.4$ where i = 1, 2.
For all the other jets, $\Delta\phi (j, \cancel{\vec E}_T) > 0.2$.
\item[A4:] We demand a hard cut on the effective mass variable, $M_{Eff} > 800$ GeV. 
\item[A5:] We put the bound on the missing enrgy $\cancel{E}_{T} > 160$ GeV. 

$M_{Eff}$ and $\cancel {E_T}$ are the two most effective cuts to reduce SM background 
events for multi-jet analyses. As shown in Fig.~\ref{histo}, these variables clearly 
separates the signal kinematical region from most of the dominant backgrounds quite 
effectively and can reduce the backgrounds in a significant amount. Most importantly, these 
cuts along with A1 and A2, reduces the large $QCD$ background to a neglible amount. 
\end{enumerate}
\subsubsection*{\bf Results}
In Table~\ref{tab1} and \ref{tab2}, we have shown numerical results of our collider analysis 
in production channels $\rho_2^0\rho^{\pm}$ and $\rho^+\rho^-$ 
respectively corresponding to the five choosen benchmark points
(as shown in Table \ref{tab3}) which satisfy the present day accepted value of the DM relic density and are
safe from the different ongoing direct detection experiment. We have
studied the SM background in detail and in Table \ref{tab5} we have shown the resulting cross sections
after appyling the aforementioned cuts. Here, we have considered NLO cross section for all the
SM background processes as provided in \cite{Alwall:2014hca}. 
\def\I{i}
\begin{center}
\begin{table}[h!]
\begin{tabular}{||c|c||}
\hline
\hline
\begin{tabular}{c|c}
    \multicolumn{2}{c}{Signal at 13 TeV}\\ 
    \hline
    BP & Cross-section (pb) \\ 
    \hline
    BP1 & 6.757\\ 
    \hline
    BP2 & 2.279 \\
    \hline
    BP3 & 1.052\\
    \hline
    BP4 & 1.296\\ 
    \hline
    BP5 & 0.760 \\  
\end{tabular}
&
\begin{tabular}{c|c|c|c|c}
    \multicolumn{5}{c}{Effective Cross section after applying cuts (fb)}\\   
    \hline 
    A0\,+\,A1 & \,\,~A2~\,\, &\,\, A3\,\, &\,\, A4\,\, &\,\, A5 \\
    \hline
    1005.05 & 175.08 & 138.45 & 22.02 & 19.15  \\
    \hline
    385.22 & 69.16 & 56.51 & 11.87 & 10.85  \\
    \hline
    189.71 & 34.63 & 29.19 & 7.36 & 6.82   \\
    \hline
    1047.86 & 145.67 & 116.94 & 14.19 & 9.82  \\
    \hline
     616.00 & 89.60 & 72.63 & 9.80 & 7.40  \\
\end{tabular}\\
\hline
\hline
\end{tabular}
\caption{Cut-flow table for the obtained signal cross section at 13 TeV LHC corresponding to $\rho^0_2 \rho^{\pm}$ channel. The five benchmark points are referred as BP1-BP5. See the text for the details of the cuts A0-A5.}
\label{tab1}
\end{table}
\end{center}
\def\I{i}
\begin{center}
\begin{table}[h!]
\begin{tabular}{||c|c||}
\hline
\hline
\begin{tabular}{c|c}
    \multicolumn{2}{c}{Signal at 13 TeV}\\
    \hline
    BP & Cross-section (pb) \\
    \hline
    BP1 & 3.419\\
    \hline
    BP2 & 1.156\\
    \hline
    BP3 & 0.532 \\
    \hline
    BP4 & 0.652 \\
    \hline
    BP5 & 0.380 \\
\end{tabular}
&
\begin{tabular}{c|c|c|c|c|c}
    \multicolumn{5}{c}{Effective Cross section after applying cuts (fb)}\\
    \hline
    A0\,+\,A1 & \,\,~A2~\,\, &\,\, A3\,\, &\,\, A4\,\, &\,\, A5 \\
    \hline
    2639.30 & 74.36 & 59.18 & 8.54 & 7.31   \\
    \hline
    880.60 & 28.77 & 23.87 & 4.95 & 4.43   \\
    \hline
    402.24 & 14.80 & 12.62 & 3.18 & 2.95  \\
    \hline
    446.80 & 63.99 & 45.54 & 5.72 & 3.76  \\
    \hline
    258.55 & 34.40 & 28.07 & 3.99 & 3.08  \\
\end{tabular}\\
\hline
\hline
\end{tabular}
\caption{Cut-flow table for the obtained signal cross section at 13 TeV LHC corresponding to $\rho^{+} \rho^{-}$ channel. The five benchmark points are referred as BP1-BP5. See the text for the details of the cuts A0-A5. }
\label{tab2}
\end{table}
\end{center}
\def\I{i}
\begin{center}
\begin{table}[h!]
\begin{tabular}{||c|c||}
\hline
\hline
\begin{tabular}{c|c}
    \multicolumn{2}{c}{SM Backgrounds at 13 TeV}\\ 
    \hline
    Channels & Cross-section (pb) \\ 
    \hline
    Z $+\leq$ 4 jets & 5.7$\times 10^4$ \\ 
    \hline
    W$^{\pm}$ + $\leq$ 4 jets & 1.9$\times 10^5$  \\
    \hline
    QCD ($\leq$ 4 jets) & 2.0$\times 10^{8}$\\
    \hline
    t\,$\bar{\rm t} + $ $\leq$ 2 jets & 722.94\\
    \hline
    W$^{\pm}$Z + $\leq$ 2 jets & 51.10 \\
    \hline
    Z\,Z +$\leq$ 2 jets & 13.71\\ 
    \hline
    Total Backgrounds & ~\\ 
\end{tabular}
&
\begin{tabular}{c|c|c|c|c}
    \multicolumn{5}{c}{Effective Cross section after applying cuts (pb)}\\   
    \hline 
    A0\,+\,A1 & \,\,~A2~\,\, &\,\, A3\,\, &\,\, A4\,\, &\,\, A5  \\
    \hline
    5.5 $\times 10^{3}$ & 361.90 & 241.60 & 11.40 & 2.20  \\
    \hline
    9.1 $\times 10^{3}$ & 783.20 & 504.00 & 18.90 & 1.50  \\
    \hline
    1.5 $\times 10^{7}$ & 3.5 $\times 10^{5}$ & 2.4 $\times 10^{5}$ & 2.5 $\times 10^{3}$ & -  \\
    
    \hline
    493.73 & 171.46 & 120.63 & 13.89 & 1.94 \\
    \hline
    19.66 & 5.37 & 3.59 & 0.50 & 0.12 \\
    \hline
    4.99 & 0.80 & 0.53 & 0.06 & 0.02  \\
    \hline
    ~ & ~ & ~ & ~ & 5.78 \\
\end{tabular}\\
\hline
\hline
\end{tabular}
\caption{Cut-flow table for the obtained cross-sections corresponding to the
relevant SM background channels for the cuts A0-A5 as mentioned in the text 
at the LHC with 13 TeV center-of-mass energy.}
\label{tab5}
\end{table}
\end{center}
In order to compute statistical significance ($\mathcal{S}$) of our signal for the different benchmark points
over the SM background we have used 
\begin{eqnarray}
\mathcal{S} = \sqrt{2 \times \left[ (s + b)\,{\rm ln}\left( 1 + \frac{s}{b}\right) -s \right]}.
\end{eqnarray}
where $s$ is the number of signal events and $b$ that of the total SM background contribution. 
In Table \ref{tab6}, we have shown the statistical significance obtained for 100 ${\rm fb}^{-1}$ 
integrated luminosity ($\mathcal L$). In the last column we have also shown the required ${\mathcal L}$ 
to achieve $3\sigma$ statistical significance for our benchmark points at 13 TeV LHC. 
\def\I{i}
\begin{center}
\begin{table}[h!]
\begin{tabular}{||c|c|c||}
\hline
\hline
\begin{tabular}{c|c}
    \multicolumn{2}{c}{Signal at 13 TeV}\\ 
    \hline
    BP & DM mass [GeV]\\ 
    \hline
    BP1 & 87.6\\ 
    \hline
    BP2 & 132.0 \\
    \hline
    BP3 & 171.1\\
    \hline
    BP4 & 86.7 \\ 
    \hline
    BP5 & 119.0 \\ 
\end{tabular}
&
\begin{tabular}{c}
    Statitical Significance ($\mathcal{S}$)\\   
    \hline 
    $\mathcal{L} = 100$ ${\rm fb^{-1}}$  \\
    \hline
     3.5\\
    \hline
     2.0\\
    \hline
     1.3\\
    \hline
     1.8\\
     \hline
     1.4 \\
\end{tabular}
&
\begin{tabular}{c}
    Required Luminosity $\mathcal{L}$ (${\rm fb^{-1}}$)\\   
    \hline 
    $\mathcal{S} = 3\sigma$ \\
    \hline
     74.4\\ 
    \hline
     223.0\\ 
    \hline
     545.3\\ 
     \hline
     282.3 \\  
    \hline
     473.9 \\ 
\end{tabular}\\
\hline
\hline
\end{tabular}
\caption{Statistical significance of the multi-jet signal corresponding to different benchmark 
points for ${\mathcal L}=100~{\rm fb}^{-1}$ integrated luminosity along with the required luminosity 
to achieve $3\sigma$ statistical significance at 13 TeV run of the LHC.}
\label{tab6}
\end{table}
\end{center}

As evident from Table~\ref{tab1}, \ref{tab2}, \ref{tab5} and \ref{tab6}, the used kinematical 
cuts are efficient enough to reduce the SM background contributions to the multi-jet channel. 
At the same time sufficient number of signal events survive leading to discovery potential 
of such a scenario at the 13 TeV run of the LHC with realistic integrated luminosities. 
The cuts A2, A4 and A5 are particularly useful in reducing the dominant background constributions 
arising from $W$ + jets, $Z$ + jets and $t\bar t$ + jets. A combination of cuts A2-A5 has reduced the 
QCD contribution to a negligible amount. As the numbers indicate in Table~\ref{tab6}, BP1 can 
be probed at the 13 TeV run of the LHC with 3$\sigma$ statistical significance with relatively 
low luminosity owing to the large production cross-section. As expected, the signal significance 
declines as the mass of $\rho_2^0$ ($\rho^{\pm}$) is increased while its mass gap 
with $\rho_1^0$ is kept same as represented by the numbers corresponding to the two subsequent 
benchmark points (BP2 and BP3). The last two benchmark points, BP4 and BP5 represent the scenario 
when the parent particles have masses significantly higher than the DM candidate. As a result, 
one would expect the cut efficiencies to improve for these benchmark points. This is reflected 
for example in the case of BP5 which has a signal significance very similar to BP3 in spite of 
having the smallest production cross-section. It can be inferred from our analysis that $\rho_2^0$ ($\rho^{\pm}$) 
masses $\sim$ 250 GeV can easily be probed at the 13 TeV LHC with a reasonable luminosity. 

\section{Conclusion and Summary}
\label{conclusion}

For the WIMP-type DM, its relic density, detection at direct detection experiments, and 
detection at collider experiments are intimately inter-related. In this work we have 
proposed a fermion DM model that can successfully explain the DM relic density, 
can be tested in future direct detection experiments, and can be produced and 
tested at the 13 TeV  run of the LHC. 

The model we propose extends the SM particle content by a SM triplet fermion
and a SM singlet fermion, as well as by a SM triplet scalar. Both new fermions are 
given Majorana masses. 
An overall discrete  $\mathbb{Z}_{2}$ symmetry is imposed and 
the corresponding charges of the particles under this symmetry is arranged 
in such as way that the only Yukawa coupling involving the new fermions is the 
one which includes the triplet fermion, the singlet fermion and the triplet 
scalar. This gives rise to the mixing between the 
neutral component of 
the SM triplet fermion mixing with the SM singlet fermion. 
The lighter of the two mass eigenstates becomes the DM candidate in the model, 
stabilised by the $\mathbb{Z}_{2}$ symmetry. 
There is also mixing  between 
the neutral as well as charged scalar degrees of freedom 
belonging to the SM doublet and triplet representations. Finally, we get two physical neutral 
scalars - one SM-like Higgs $h_1$ of mass 125.5 GeV and another heavier BSM Higgs $h_2$ 
whose 
mass we keep as a free parameter in the model. From the charged scalar sector, we get 
physical charged scalars $H^\pm$ while the other degree of freedom becomes the charged 
Goldstone boson, which is `eaten up' to give mass to the $W^{\pm}$ bosons. 
The presence of the triplet scalar as well as the mixing between the triplet and 
singlet fermions lead to additional $s$-channel diagrams mediated by $h_1$ and $h_2$ 
as well as $t$-channel diagrams mediated by the new fermions $\rho^0_{1,2}$ and 
$\rho^{\pm}$ with $H^\pm$ or $h_2$ in their final states. These additional diagrams 
allow for resonant production of DM at (1) $h_1$ mediated $s$-channel processes, 
(2) $h_2$ mediated $s$-channel processes, and (3) $t$-channel diagrams 
with $H^\pm$ or $h_2$ in their final states. This allows us to satisfy the 
observed DM relic density by Planck with 
DM masses in the 100 GeV range. We study the impact of the 
model parameters on the DM relic density. We also study the possibility of testing 
this model at the current and future direct detection experiments,  Xenon 1T 
and Darwin. 

Finally we study  the LHC phenomenology for few benchmark points (BP) and show that 
this model is testable in the very near future run of  LHC. 
The model proposed in \cite{Ma:2008cu} had only the triplet fermion 
(and an inert doublet scalar) where the neutral component of the triplet becomes the 
DM stabilised by the $\mathbb{Z}_{2}$ symmetry. 
The relic density of the fermionic DM in that model was governed by the $t$-channel 
processes involving only the $W^\pm$, SM-like Higgs and SM fermions. As a result, the 
DM mass was seen to be 2.37 TeV for the Planck bound to be satisfied. Hence, this 
model cannot be tested at the LHC. Since the DM mass in our model is 
${\cal O} (100$ GeV), 
therefore, we can produce them in the collider at the 13 TeV run of LHC with a reasonably
large cross-section. In this work we analysed the multi jets + missing energy
signal. We also considered the low mass difference between the DM ($\rho_{1}^{0}$)
and the next-to-lightest neutral particle ($\rho_{2}^{0}$), that might lead to softer jets. However, high $p_T$ jets may come from the ISR.
Corresponding to this signal we figured out the  dominant SM backgrounds
for multi jets + missing energy signal. By suitably choosing the cuts we have reduced the
SM backgrounds and simultaneously increased the  statistical significance of the signal. 
We showed that for 100 fb$^{-1}$, we could observe the DM with $3.5\sigma$ statistical 
significance for one of the BP. The  
statistical significance was seen reduce as the mass of $\rho^0_2$ and $\rho^\pm$ 
was increased. We also studied how much luminosity would be 
required to probe this model with a $3\sigma$ statistical significance for our BPs. 

A final comment on our model is in order. While we have focussed only on 
explaining the DM relic density within the context of the present model, 
we can easily extend it to generate the neutrino masses by Type I seesaw
mechanism. This can be done by introducing right-handed neutrinos. 

In conclusion, the present model allows for low mass fermionic DM that satisfactorily  
produces the observed the relic density of the universe. It can be tested 
at the current and next-generation DM direct detection experiments. More 
importantly the 100 GeV mass range of the DM candidate in this model allows 
its production and detection at the LHC. The 13 TeV LHC can discover this fermionic 
DM candidate for with more than $3\sigma$ statistical significance with reasonable luminosity. 



\acknowledgments
We acknowledge the HRI cluster computing facility (http://cluster.hri.res.in).
The authors would like to thank the Department of Atomic Energy (DAE) Neutrino Project of Harish-Chandra Research Institute. This project has received funding from the European Union's Horizon 2020 research and innovation programme InvisiblesPlus RISE under the Marie Sklodowska-Curie grant agreement No 690575. This project has received funding from the European Union's Horizon 2020 research and innovation programme Elusives ITN under the Marie Sklodowska- Curie grant agreement No 674896. 
SK would also like to thank IOP, Bhubaneswar for hospitality during the
initial stage of this work.
MM acknowledges the support of DST-INSPIRE FACULTY research grant.   
SM would also like to thank HRI and RECAPP, HRI for hospitality and financial support during the initial phase of this work.
	
\section*{Appendix}
\label{App:AppendixA}
The factors $I_4^j$ and $I_3^i$ take the following form \cite{Bergstrom:1997fh, Bern:1997ng},
as used in the Eq.\,(\ref{amplitude-gamma-rays}), for the $A_{W\rho}$ part $I_4^j$ has the
following structure,
\begin{eqnarray}
I_4^a &= & \frac{I_3^2(M_{W},M_{\rho^{\pm}}) I_{2}^1(M_{W})}
{M_{\rho_1^0}^2 + M_{\rho^{\pm}}^2 - M_{W}^2}, \nn \\
I_4^b &=& \frac{I_3^2(M_{\rho^{\pm}},M_{W}) - I_3^1(M_{\rho^{\pm}})}
{M_{\rho_1^0}^2 - M_{\rho^{\pm}}^2 + M_{W}^2}, \nn \\
I_4^c &=& \frac{I_3^2(M_{\rho^{\pm}}, M_{W}) - I_3^2(M_{W},M_{\rho^{\pm}})}
{M_{\rho^{\pm}}^2 - M_{W}^2}
\end{eqnarray}	
and for determing the amplitude $A_{H\rho}$, we just need to
replace the $W^{\pm}$ gauge boson mass ($M_{W}$) by the mass ($M_{H^{\pm}}$) of
the charged scalar $H^{\pm}$. Now the $I_3^i$ components take the following form,

\begin{eqnarray}
I_3^a(m) = \Bigg\{ \begin{matrix} \frac{1}{8\,M^2_{\rho_1^0}} 
[ \log^2(\frac{1+x}{1-x}) - \pi^2 - 2 i \pi \log(\frac{1+x}{1-x})], & m \leq M_{\rho_1^0}  
\\ - \frac{1}{2 M_{\rho_1^0}^2}\left(\tan^{-1}(\frac{1}{m^2/M_{\rho_1^0}^2}-1) \right),&
m > M_{\rho_1^0} \end{matrix} 
\end{eqnarray}

and the other component takes the form,
\begin{eqnarray}
I_3^b(m_1,m_2) &=& \frac{1}{2 M_{\rho_1^0}^2}
\left[ Li_{2}\left(\frac{m_1^2 - M_{\rho_1^0}^2 - m_2^2 - \sqrt{\Delta_1}}{2 m_1^2}\right) 
+ Li_{2}\left(\frac{m_1^2 - M_{\rho_1^0}^2 - m_2^2 + \sqrt{\Delta_1}}{2 m_1^2}\right)\right]
\nn \\
&& - \frac{1}{2 M_{\rho_1^0}^2}
\left[ Li_{2}\left(\frac{m_1^2 + M_{\rho_1^0}^2 - m_2^2 - \sqrt{\Delta_2}}{2 m_1^2}\right) 
+ Li_{2}\left(\frac{m_1^2 + M_{\rho_1^0}^2 - m_2^2 + \sqrt{\Delta_2}}{2 m_1^2}\right)\right] \nn \\ 
\end{eqnarray}
where
\begin{eqnarray}
x &=& \sqrt{1 - m^2/M_{\rho_1^0}} \nn \\
\Delta_1 &=& (m_1^2 + M_{\rho_1^0}^2 - m_2^2)^2 + 4 M_{\rho_1^0}^2 m_2^2 \nn \\
\Delta_1 &=& (m_1^2 - M_{\rho_1^0}^2 - m_2^2)^2 - 4 M_{\rho_1^0}^2 m_2^2 \nn
\end{eqnarray}


\begin{thebibliography}{99}
\bibitem{Sofue:2000jx} 
Y.~Sofue and V.~Rubin,
``{\it Rotation curves of spiral galaxies}'',
Ann.\ Rev.\ Astron.\ Astrophys.\  {\bf 39}, 137 (2001)
[astro-ph/0010594].

\bibitem{Bartelmann:1999yn} 
M.~Bartelmann and P.~Schneider,
``{\it Weak gravitational lensing}'',
Phys.\ Rept.\  {\bf 340}, 291 (2001)
[astro-ph/9912508].

\bibitem{Clowe:2003tk} 
D.~Clowe, A.~Gonzalez and M.~Markevitch,
``{\it Weak lensing mass reconstruction of the interacting
cluster 1E0657-558: Direct evidence for the existence of dark matter}'',
Astrophys.\ J.\  {\bf 604}, 596 (2004)
[astro-ph/0312273].



\bibitem{Harvey:2015hha} 
D.~Harvey, R.~Massey, T.~Kitching, A.~Taylor and E.~Tittley,
``{\it The non-gravitational interactions of dark matter
in colliding galaxy clusters}'',
Science {\bf 347}, 1462 (2015)
[arXiv:1503.07675 [astro-ph.CO]].

\bibitem{Hinshaw:2012aka} 
G.~Hinshaw {\it et al.} [WMAP Collaboration],
``{\it Nine-Year Wilkinson Microwave Anisotropy Probe (WMAP)
Observations: Cosmological Parameter Results}'',
Astrophys.\ J.\ Suppl.\  {\bf 208}, 19 (2013)
[arXiv:1212.5226 [astro-ph.CO]].

\bibitem{Ade:2015xua} 
P.~A.~R.~Ade {\it et al.} [Planck Collaboration],
``{\it Planck 2015 results. XIII. Cosmological parameters}'',
arXiv:1502.01589 [astro-ph.CO].

\bibitem{Hisano:2002fk} 
  J.~Hisano, S.~Matsumoto and M.~M.~Nojiri,
  ``{\it Unitarity and higher order corrections in neutralino dark matter annihilation into two photons}'',
  Phys.\ Rev.\ D {\bf 67}, 075014 (2003)
  [hep-ph/0212022].

\bibitem{Hisano:2003ec} 
  J.~Hisano, S.~Matsumoto and M.~M.~Nojiri,
  ``{\it Explosive dark matter annihilation}'',
  Phys.\ Rev.\ Lett.\  {\bf 92}, 031303 (2004)
  [hep-ph/0307216].



\bibitem{Hisano:2004ds} 
  J.~Hisano, S.~Matsumoto, M.~M.~Nojiri and O.~Saito,
  ``{\it Non-perturbative effect on dark matter annihilation and gamma ray signature from galactic center}'',
  Phys.\ Rev.\ D {\bf 71}, 063528 (2005)
  [hep-ph/0412403].


\bibitem{Cirelli:2005uq} 
  M.~Cirelli, N.~Fornengo and A.~Strumia,
  ``{\it Minimal dark matter}'',
  Nucl.\ Phys.\ B {\bf 753}, 178 (2006)
  [hep-ph/0512090].



\bibitem{Hisano:2006nn} 
  J.~Hisano, S.~Matsumoto, M.~Nagai, O.~Saito and M.~Senami,
  ``{\it Non-perturbative effect on thermal relic abundance of dark matter}'',
  Phys.\ Lett.\ B {\bf 646}, 34 (2007)
  [hep-ph/0610249].

\bibitem{Ma:2008cu} 
  E.~Ma and D.~Suematsu,
  ``{\it Fermion Triplet Dark Matter and Radiative Neutrino Mass}'',
  Mod.\ Phys.\ Lett.\ A {\bf 24}, 583 (2009)
  [arXiv:0809.0942 [hep-ph]].

\bibitem{Cohen:2013ama} 
  T.~Cohen, M.~Lisanti, A.~Pierce and T.~R.~Slatyer,
  ``{\it Wino Dark Matter Under Siege}'',
  JCAP {\bf 1310}, 061 (2013)
  [arXiv:1307.4082 [hep-ph]].


\bibitem{Chardonnet:1993wd} 
  P.~Chardonnet, P.~Salati and P.~Fayet,
  ``{\it Heavy triplet neutrinos as a new dark matter option}'',
  Nucl.\ Phys.\ B {\bf 394}, 35 (1993).

\bibitem{Biggio:2011ja} 
  C.~Biggio and F.~Bonnet,
  ``{\it Implementation of the Type III Seesaw Model in FeynRules/MadGraph and Prospects for Discovery with Early LHC Data}'',
  Eur.\ Phys.\ J.\ C {\bf 72}, 1899 (2012)
  [arXiv:1107.3463 [hep-ph]].


\bibitem{Hirsch:2013ola} 
  M.~Hirsch, R.~A.~Lineros, S.~Morisi, J.~Palacio, N.~Rojas and J.~W.~F.~Valle,
  ``{\it WIMP dark matter as radiative neutrino mass messenger}'',
  JHEP {\bf 1310}, 149 (2013)
  [arXiv:1307.8134 [hep-ph]].

\bibitem{Chaudhuri:2015pna} 
  A.~Chaudhuri, N.~Khan, B.~Mukhopadhyaya and S.~Rakshit,
  Phys.\ Rev.\ D {\bf 91}, 055024 (2015)
  doi:10.1103/PhysRevD.91.055024
  [arXiv:1501.05885 [hep-ph]].

\bibitem{Bhattacharya:2017sml} 
  S.~Bhattacharya, N.~Sahoo and N.~Sahu,
  ``{\it Singlet-Doublet Fermionic Dark Matter, Neutrino Mass and Collider Signatures}'',
  Phys.\ Rev.\ D {\bf 96}, no. 3, 035010 (2017)
  [arXiv:1704.03417 [hep-ph]].


\bibitem{Frigerio:2009wf} 
  M.~Frigerio and T.~Hambye,
  ``{\it Dark matter stability and unification without supersymmetry}'',
  Phys.\ Rev.\ D {\bf 81}, 075002 (2010)
  [arXiv:0912.1545 [hep-ph]].
  
\bibitem{Mambrini:2016dca} 
  Y.~Mambrini, N.~Nagata, K.~A.~Olive and J.~Zheng,
  ``{\it Vacuum Stability and Radiative Electroweak Symmetry Breaking in an SO(10) Dark Matter Model}'',
  Phys.\ Rev.\ D {\bf 93}, no. 11, 111703 (2016)
  [arXiv:1602.05583 [hep-ph]].


\bibitem{Erler:2004nh} 
  J.~Erler and P.~Langacker,
  ``{\it Electroweak model and constraints on new physics}'',
  hep-ph/0407097.

\bibitem{Erler:2008ek} 
  J.~Erler and P.~Langacker,
  ``{\it Electroweak Physics}'',
  Acta Phys.\ Polon.\ B {\bf 39}, 2595 (2008)
  [arXiv:0807.3023 [hep-ph]].


\bibitem{Chen:2008jg} 
  M.~C.~Chen, S.~Dawson and C.~B.~Jackson,
  ``{\it Higgs Triplets, Decoupling, and Precision Measurements}'',
  Phys.\ Rev.\ D {\bf 78}, 093001 (2008)
  [arXiv:0809.4185 [hep-ph]].



\bibitem{Abada:2008ea} 
  A.~Abada, C.~Biggio, F.~Bonnet, M.~B.~Gavela and T.~Hambye,
  ``{\it $\mu ->$ e $\gamma$ and tau ---> l gamma decays in the fermion triplet seesaw model}'',
  Phys.\ Rev.\ D {\bf 78}, 033007 (2008)
  [arXiv:0803.0481 [hep-ph]].



\bibitem{Belanger:2008sj} 
G.~Belanger, F.~Boudjema, A.~Pukhov and A.~Semenov,
``{\it Dark matter direct detection rate in a generic model with micrOMEGAs 2.2}'',
Comput.\ Phys.\ Commun.\  {\bf 180}, 747 (2009)
[arXiv:0803.2360 [hep-ph]].

\bibitem{Akerib:2015rjg} 
  D.~S.~Akerib {\it et al.} [LUX Collaboration],
  ``{\it Improved Limits on Scattering of Weakly Interacting Massive Particles from Reanalysis of 2013 LUX Data}'',
  Phys.\ Rev.\ Lett.\  {\bf 116}, no. 16, 161301 (2016)
  [arXiv:1512.03506 [astro-ph.CO]].


\bibitem{Akerib:2016vxi} 
  D.~S.~Akerib {\it et al.} [LUX Collaboration],
  ``{\it Results from a search for dark matter in the complete LUX exposure}'',
  Phys.\ Rev.\ Lett.\  {\bf 118}, no. 2, 021303 (2017)
  [arXiv:1608.07648 [astro-ph.CO]].

\bibitem{Aprile:2015uzo} 
  E.~Aprile {\it et al.} [XENON Collaboration],
  ``{\it Physics reach of the XENON1T dark matter experiment}'',
  JCAP {\bf 1604}, no. 04, 027 (2016)
  [arXiv:1512.07501 [physics.ins-det]].
	
\bibitem{Aprile:2017iyp} 
  E.~Aprile {\it et al.} [XENON Collaboration],
  ``{\it First Dark Matter Search Results from the XENON1T Experiment}'',
  arXiv:1705.06655 [astro-ph.CO].

\bibitem{Cui:2017nnn} 
  X.~Cui {\it et al.} [PandaX-II Collaboration],
  ``{\it Dark Matter Results From 54-Ton-Day Exposure of PandaX-II Experiment}'',
  arXiv:1708.06917 [astro-ph.CO].
	
\bibitem{Aalbers:2016jon} 
  J.~Aalbers {\it et al.} [DARWIN Collaboration],
  ``{\it DARWIN: towards the ultimate dark matter detector}'',
  JCAP {\bf 1611}, 017 (2016)
  [arXiv:1606.07001 [astro-ph.IM]].


\bibitem{Abramowski:2013ax} 
  A.~Abramowski {\it et al.} [H.E.S.S. Collaboration],
  ``{\it Search for Photon-Linelike Signatures from Dark Matter Annihilations with 
  H.E.S.S.}'',
  Phys.\ Rev.\ Lett.\  {\bf 110}, 041301 (2013)
  [arXiv:1301.1173 [astro-ph.HE]].

\bibitem{Profumo:2016idl} 
  S.~Profumo, F.~S.~Queiroz and C.~E.~Yaguna,
  ``{\it Extending Fermi-LAT and H.E.S.S. Limits on Gamma-ray Lines from Dark Matter Annihilation}'',
  Mon.\ Not.\ Roy.\ Astron.\ Soc.\  {\bf 461}, no. 4, 3976 (2016)
  [arXiv:1602.08501 [astro-ph.HE]].


\bibitem{Ackermann:2013uma} 
  M.~Ackermann {\it et al.} [Fermi-LAT Collaboration],
  ``{\it Search for Gamma-ray Spectral Lines with the Fermi Large Area Telescope and Dark Matter Implications}'',
  Phys.\ Rev.\ D {\bf 88}, 082002 (2013)
  [arXiv:1305.5597 [astro-ph.HE]].

\bibitem{Ackermann:2015lka} 
  M.~Ackermann {\it et al.} [Fermi-LAT Collaboration],
  ``{\it Updated search for spectral lines from Galactic dark matter interactions with pass 8 data from the Fermi Large Area Telescope}'',
  Phys.\ Rev.\ D {\bf 91}, no. 12, 122002 (2015)
  [arXiv:1506.00013 [astro-ph.HE]].

\bibitem{Bergstrom:1997fh} 
  L.~Bergstrom and P.~Ullio,
  ``{\it Full one loop calculation of neutralino annihilation into two photons}'',
  Nucl.\ Phys.\ B {\bf 504}, 27 (1997)
  [hep-ph/9706232].


\bibitem{Bern:1997ng} 
  Z.~Bern, P.~Gondolo and M.~Perelstein,
  ``{\it Neutralino annihilation into two photons}'',
  Phys.\ Lett.\ B {\bf 411}, 86 (1997)
  [hep-ph/9706538].

\bibitem{Khachatryan:2016vau} 
  G.~Aad {\it et al.} [ATLAS and CMS Collaborations],
  ``{\it Measurements of the Higgs boson production and decay rates and constraints on its couplings from a combined ATLAS and CMS analysis of the LHC pp collision data at $ \sqrt{s}=7 $ and 8 TeV}'',
  JHEP {\bf 1608}, 045 (2016)
  [arXiv:1606.02266 [hep-ex]].
	

\bibitem{Alloul:2013bka} 
  A.~Alloul, N.~D.~Christensen, C.~Degrande, C.~Duhr and B.~Fuks,
  ``{\it FeynRules  2.0 - A complete toolbox for tree-level phenomenology}'',
  Comput.\ Phys.\ Commun.\  {\bf 185}, 2250 (2014)
  [arXiv:1310.1921 [hep-ph]].


\bibitem{Belanger:2013oya} 
  G.~Belanger, F.~Boudjema, A.~Pukhov and A.~Semenov,
  ``{\it micrOMEGAs-3: A program for calculating dark matter observables}'',
  Comput.\ Phys.\ Commun.\  {\bf 185}, 960 (2014)
  [arXiv:1305.0237 [hep-ph]].


\bibitem{Alves:2010za} 
  D.~S.~M.~Alves, E.~Izaguirre and J.~G.~Wacker,
  ``{\it It's On: Early Interpretations of ATLAS Results in Jets and Missing Energy Searches}'',
  Phys.\ Lett.\ B {\bf 702}, 64 (2011)
  [arXiv:1008.0407 [hep-ph]].


\bibitem{Aaboud:2017gsl} 
  M.~Aaboud {\it et al.} [ATLAS Collaboration],
  ``{\it Search for heavy resonances decaying into $WW$ in the $e\nu\mu\nu$ final state in $pp$ collisions at $\sqrt{s}=13$ TeV with the ATLAS detector}'',
  Eur.\ Phys.\ J.\ C {\bf 78}, no. 1, 24 (2018)
  [arXiv:1710.01123 [hep-ex]]. 
\bibitem{Aaboud:2017rel} 
  M.~Aaboud {\it et al.} [ATLAS Collaboration],
  ``{\it Search for heavy $ZZ$ resonances in the $\ell^+\ell^-\ell^+\ell^-$ and $\ell^+\ell^-\nu\bar\nu$ final states using proton proton collisions at $\sqrt{s}= 13$ TeV with the ATLAS detector}'',
  arXiv:1712.06386 [hep-ex].
\bibitem{Aaboud:2017yyg} 
  M.~Aaboud {\it et al.} [ATLAS Collaboration],
  ``{\it Search for new phenomena in high-mass diphoton final states using 37 fb$^{-1}$ of proton--proton collisions collected at $\sqrt{s}=13$ TeV with the ATLAS detector}'',
  Phys.\ Lett.\ B {\bf 775}, 105 (2017)
  [arXiv:1707.04147 [hep-ex]].  	
\bibitem{Khachatryan:2015qxa} 
  V.~Khachatryan {\it et al.} [CMS Collaboration],
  ``{\it Search for a charged Higgs boson in pp collisions at $ \sqrt{s}=8 $ TeV}'',
  JHEP {\bf 1511}, 018 (2015)
  [arXiv:1508.07774 [hep-ex]].
\bibitem{Khachatryan:2015uua} 
  V.~Khachatryan {\it et al.} [CMS Collaboration],
  ``{\it Search for a light charged Higgs boson decaying to $ \mathrm{c}\overline{\mathrm{s}} $ in pp collisions at $ \sqrt{s}=8 $ TeV}'',
  JHEP {\bf 1512}, 178 (2015)
  [arXiv:1510.04252 [hep-ex]].
\bibitem{CMS:2016qoa} 
  CMS Collaboration [CMS Collaboration],
  ``{\it Search for Charged Higgs boson to ${\rm c\bar{b}}$ in lepton+jets channel using top quark pair events}'',
  CMS-PAS-HIG-16-030.
\bibitem{ATLAS:2016qiq} 
  The ATLAS collaboration [ATLAS Collaboration],
  ``{\it Search for charged Higgs bosons in the $H^{\pm}\to tb$ decay channel in $pp$ collisions at $\sqrt{s}=13$ TeV using the ATLAS detector}'',
  ATLAS-CONF-2016-089.
\bibitem{ATLAS:2017uun} 
  The ATLAS collaboration [ATLAS Collaboration],
  ``{\it Search for electroweak production of supersymmetric particles in the two and three lepton final state at $\boldmath{\sqrt{s}=13\,}$TeV with the ATLAS detector}'',
  ATLAS-CONF-2017-039.
\bibitem{Aaboud:2017nhr} 
  M.~Aaboud {\it et al.} [ATLAS Collaboration],
  ``{\it Search for the direct production of charginos and neutralinos  in final states with tau leptons in $\sqrt{s} = $ 13 TeV $pp$ collisions with the ATLAS detector}'',
  Eur.\ Phys.\ J C {\bf 78}, 154 (2018)
  [arXiv:1708.07875 [hep-ex]].



\bibitem{Mangano:2006rw} 
  M.~L.~Mangano, M.~Moretti, F.~Piccinini and M.~Treccani,
  ``{\it Matching matrix elements and shower evolution for top-quark production in hadronic collisions}'',
  JHEP {\bf 0701}, 013 (2007)
  [hep-ph/0611129].

\bibitem{Hoche:2006ph} 
  S.~Hoeche, F.~Krauss, N.~Lavesson, L.~Lonnblad, M.~Mangano, A.~Schalicke and S.~Schumann,
  ``{\it Matching parton showers and matrix elements}'',
  hep-ph/0602031.

\bibitem{Alwall:2014hca} 
  J.~Alwall {\it et al.},
  ``{\it The automated computation of tree-level and next-to-leading order differential cross sections, and their matching to parton shower simulations}'',
  JHEP {\bf 1407}, 079 (2014)
  [arXiv:1405.0301 [hep-ph]].

\bibitem{Alwall:2011uj} 
  J.~Alwall, M.~Herquet, F.~Maltoni, O.~Mattelaer and T.~Stelzer,
  ``{\it MadGraph 5 : Going Beyond}'',
  JHEP {\bf 1106}, 128 (2011)
  [arXiv:1106.0522 [hep-ph]].

\bibitem{Pumplin:2002vw} 
  J.~Pumplin, D.~R.~Stump, J.~Huston, H.~L.~Lai, P.~M.~Nadolsky and W.~K.~Tung,
  ``{\it New generation of parton distributions with uncertainties from global QCD analysis}'',
  JHEP {\bf 0207}, 012 (2002)
  [hep-ph/0201195].

\bibitem{Sjostrand:2006za} 
  T.~Sjostrand, S.~Mrenna and P.~Z.~Skands,
  ``{\it PYTHIA 6.4 Physics and Manual}'',
  JHEP {\bf 0605}, 026 (2006)
  [hep-ph/0603175].


\bibitem{deFavereau:2013fsa} 
  J.~de Favereau {\it et al.} [DELPHES 3 Collaboration],
  ``{\it DELPHES 3, A modular framework for fast simulation of a generic collider experiment}'',
  JHEP {\bf 1402}, 057 (2014)
  [arXiv:1307.6346 [hep-ex]].


\bibitem{Selvaggi:2014mya} 
  M.~Selvaggi,
  ``{\it DELPHES 3: A modular framework for fast-simulation of generic collider experiments}'',
  J.\ Phys.\ Conf.\ Ser.\  {\bf 523}, 012033 (2014).
  
\bibitem{Mertens:2015kba} 
  A.~Mertens,
  ``{\it New features in Delphes 3}'',
  J.\ Phys.\ Conf.\ Ser.\  {\bf 608}, no. 1, 012045 (2015).

\bibitem{Cacciari:2008gp} 
  M.~Cacciari, G.~P.~Salam and G.~Soyez,
  ``{\it The Anti-k(t) jet clustering algorithm}'',
  JHEP {\bf 0804}, 063 (2008)
  [arXiv:0802.1189 [hep-ph]].
 
\bibitem{Cacciari:2011ma} 
  M.~Cacciari, G.~P.~Salam and G.~Soyez,
  ``{\it FastJet User Manual}'',
  Eur.\ Phys.\ J.\ C {\bf 72}, 1896 (2012)
  [arXiv:1111.6097 [hep-ph]].
 
 
\bibitem{Aad:2014wea} 
  G.~Aad {\it et al.} [ATLAS Collaboration],
  ``{\it Search for squarks and gluinos with the ATLAS detector in final states with jets and missing transverse momentum using $\sqrt{s}=8$ TeV proton--proton collision data}'',
  JHEP {\bf 1409}, 176 (2014)
  [arXiv:1405.7875 [hep-ex]].

\bibitem{Dutta:2015exw} 
  J.~Dutta, P.~Konar, S.~Mondal, B.~Mukhopadhyaya and S.~K.~Rai,
  ``{\it A Revisit to a Compressed Supersymmetric Spectrum with 125 GeV Higgs}'',
  JHEP {\bf 1601}, 051 (2016)
  [arXiv:1511.09284 [hep-ph]].

  	
\end{thebibliography}
\end{document}